\documentclass[a4paper,fleqn]{cas-sc}

\usepackage[numbers]{natbib}

\usepackage{cite}
\usepackage{amsmath,amssymb,amsfonts}
\usepackage{graphicx}
\usepackage{textcomp}

\usepackage{multirow}
\usepackage{float}
\usepackage{balance}
\usepackage{booktabs}
\usepackage{soul}
\usepackage{adjustbox}
\usepackage{subcaption}
\usepackage{comment}
\usepackage[inline]{enumitem}
\usepackage{makecell}
\usepackage{wrapfig}
\usepackage{lscape}

\newcommand{\subr}[1]{{\small\texttt{r/#1}}}

\newcommand{\rev}[1]{\textcolor{black}{#1}}

\def\tsc#1{\csdef{#1}{\textsc{\lowercase{#1}}\xspace}}
\tsc{WGM}
\tsc{QE}

\begin{document}
\let\WriteBookmarks\relax
\def\floatpagepagefraction{1}
\def\textpagefraction{.001}

\shorttitle{Beyond Trial-and-Error: Predicting User Abandonment After a Moderation Intervention}    

\shortauthors{Tessa et al.}  

\title [mode = title]{Beyond Trial-and-Error: Predicting User Abandonment After a Moderation Intervention}  

\tnotemark[1] 

\tnotetext[1]{\textcolor{red}{Article published in \textit{Engineering Applications of Artificial Intelligence}. DOI: \href{https://doi.org/10.1016/j.engappai.2025.112375}{10.1016/j.engappai.2025.112375}. \\Please, cite the published version.}}

\author[1]{Benedetta Tessa}

\ead{benedetta.tessa@iit.cnr.it}

\author[1,2]{Lorenzo Cima}[orcid=0009-0001-5815-0437]
\ead{lorenzo.cima@iit.cnr.it}
\cormark[1]

\author[1]{Amaury Trujillo}\ead{amaury.trujillo@iit.cnr.it}

\author[2]{Marco Avvenuti}\ead{marco.avvenuti@unipi.it}

\author[1]{Stefano Cresci}\ead{stefano.cresci@iit.cnr.it}

\affiliation[1]{organization={IIT-CNR},
            addressline={Via Moruzzi 1},
            code={56124},
            city={Pisa},
            country={Italy}}

\affiliation[2]{organization={University of Pisa - Department of Information Engineering},
            addressline={Via Caruso 16}, 
            code={56122},
            city={Pisa},
            country={Italy}}

\cortext[1]{Corresponding author}

\begin{abstract}
\protect{Current content moderation follows a reactive, trial-and-error approach, where interventions are applied and their effects are only measured post-hoc. In contrast, we introduce a proactive, predictive approach that enables moderators to anticipate the impact of their actions before implementation. We propose and tackle the new task of predicting user abandonment following a moderation intervention. We study the reactions of 16,540 users to a massive ban of online communities on Reddit, training a set of binary classifiers to identify those users who would abandon the platform after the intervention---a problem of great practical relevance. We leverage a dataset of 13.8 million posts to compute a large and diverse set of 142 features, which convey information about the activity, toxicity, relations, and writing style of the users. We obtain promising results, with the best-performing model achieving \textit{micro F1-score} $= 0.914$. 
Our model shows robust generalizability when applied to users from previously unseen communities. Furthermore, we identify activity features as the most informative predictors, followed by relational and toxicity features, while writing style features exhibit limited utility. Theoretically, our results demonstrate the feasibility of adopting a predictive machine learning approach to estimate the effects of moderation interventions. Practically, this work marks a fundamental shift from reactive to predictive moderation, equipping platform administrators with intelligent tools to strategically plan interventions, minimize unintended consequences, and optimize user engagement.}
\end{abstract}


\maketitle

\section{Introduction}
\label{sec:introduction}
Online social media has greatly facilitated the sharing of information and opinions of all sorts, both locally and across the globe~\citep{dwivedi2018social}.
However, this also includes less desirable behavior, such as the spread of online toxic content, which besides potentially harming targeted users, is detrimental to the overall health and inclusivity of online environments~\citep{abarna2022identification,saumya2024filtering}.
Online platforms are thus constrained to moderate such content, not only to maintain a safe and welcoming environment, but also because of legal compliance requirements~\citep{trujillo2023dsa,tessa2025improving,shahi2025year}.
Moderation strategies can be broadly divided into two main categories: hard and soft. The former consists of permanently removing users, groups, or posts that are violating community guidelines, such as posts deemed toxic. This type of intervention ---also called \textit{deplatforming}--- can have unintended consequences because certain users perceive content removals as a form of censorship, especially if unmotivated. As a consequence, they may feel less prone to post again, or they could even decide to migrate to other more tolerant platforms tolerated~\citep{gorwa2020algorithmic,jhaver2019does,horta2021platform}. This led to the proposal of the second broad form of moderation ---soft moderation--- which warns users about the harmfulness of a content without removing it~\citep{singhal2023sok, zannettou2021won}.
The growing consensus among the thriving content moderation literature is that gradual approaches that initially apply less punitive interventions, and that subsequently escalate the severity whenever required, are generally more effective than other solutions~\citep{kiesler2012regulating}.
Nevertheless, hard interventions are still the most widely used form of content moderation. 
A well-known example of this kind is the so-called \textit{Great Ban}, a massive deplatforming intervention enforced by Reddit in response to the rise of toxic and hateful speech on the platform, which involved the permanent ban of around 2,000 subreddits in June 2020~\citep{cima2024great}. Among the banned subreddits was \subr{The\_Donald}, a community of Trump supporters that was one of the most popular political subreddits at the time~\citep{massachs2020roots}. Before the ban, the community had already faced two milder moderation interventions: a \emph{quarantine} and a \emph{restriction} that limited its visibility and the users that could serve as moderators of the community~\citep{trujillo2022make}. 
This sequence of interventions has been the subject of several studies that assessed changes in activity levels, use of language, political bias, factual news sharing, and toxicity~\citep{trujillo2022make,trujillo2023one,shen2022tale,jhaver2021evaluating,etta2022comparing}.
These works found that many users were not much affected by the interventions, showing no significant change in behavior afterward. However, others seemed to be resentful of the platform and significantly decreased their activity, so much so that some completely abandoned the platform~\citep{cima2024great,cima2025investigating}.

The previous results highlight the complexity of content moderation. First, in content moderation \textit{one size does not fit all}, and effective moderation interventions should necessarily be tailored to the targets of the moderation~\citep{cresci2022personalized,cima2025contextualized}. Furthermore, the current moderation strategy follows a \textit{trial-and-error} approach, where platforms apply sequences of interventions until they meet the desired outcome, or force the misbehaving users out of the platform. \subr{The\_Donald} was a prime example of this strategy, as it was subject to two unsuccessful interventions prior to the Great Ban~\citep{trujillo2022make}. Given that the first of such interventions was applied in June 2019, while the last in June 2020, it took several months for the platform to evaluate the effects of each intervention and to impose further restrictions. Importantly, during that time the misbehaving users continued to spread toxic content and to engage in harmful behaviors.

There is thus a pressing need to develop intelligent systems capable of supporting moderators and platform administrators by estimating the likely effect of a moderation intervention \textit{before} its application. This would open up the possibility to plan interventions ahead of time, rather than to assess and correct afterwards. To date, however, no evidence exists as to whether the effects of a moderation intervention are predictable.

\subsection{Contributions}
Motivated by the need for a strategic planning and deployment of moderation actions, we introduce and tackle the new task of predicting the effect of a moderation intervention. While many studies carried out post-hoc descriptive analyses of the effects of moderation interventions~\citep{shen2022tale,zannettou2021won,jhaver2021evaluating}, here for the first time we adopt a predictive approach. We leverage the results of previous studies on the Great Ban as our ground-truth for the effects of that intervention~\citep{trujillo2021echo,cima2024great}. Then, we analyze user behavioral traces before the Great Ban to predict possible changes in the behavior of those users after the ban.
We specifically focus on predicting those users who abandon the platform after the ban, in a binary classification task. This choice is both theoretically and practically relevant. Changes in user activity are among the most widely studied effects of past moderation interventions~\citep{trujillo2022make,trujillo2021echo,chandrasekharan2022quarantined}. As such, from the theoretical standpoint it is interesting to assess the extent to which those effects are predictable. Additionally, the task is also practically relevant since the loss of users negatively affects platform revenues, as user engagement and popularity are paramount in the economic success of social media~\citep{carlson2020you,habib2022exploring}.
 Moreover, the abandonment of influential users can also trigger chain reactions, as their followers may also choose to leave the platform, further worsening the issue.
The main results of this work are summarized in the following:
\begin{itemize}
    \item We proposed the new task of predicting users who abandon the platform following a moderation intervention and we evaluated different classification algorithms, hyper-parameters, and features sets for the task. Our best classifier is capable of detecting abandoning and non-abandoning users with a promising \textit{micro F1 score} $= 0.914$.
    \item We employed four different classes of features: those based on user (\textit{i}) activity, (\textit{ii}) toxicity, (\textit{iii}) writing style, and (\textit{iv}) relational characteristics. Activity features provided the most information to the classifiers, along with relational and toxicity ones. Instead, writing style features proved to be less informative, despite being widely used.
    \item We assessed the impact of user activity on the classification performance. We found that users exhibiting low activity are accurately classified, unlike users with high activity for which we obtained worse results. This finding suggests that more features should be designed to better characterize the online behavior of the most active users.
\end{itemize}

\subsection{Significance and Applicability}
\begin{figure}[t]
  \centering
  \includegraphics[width = 1\linewidth]{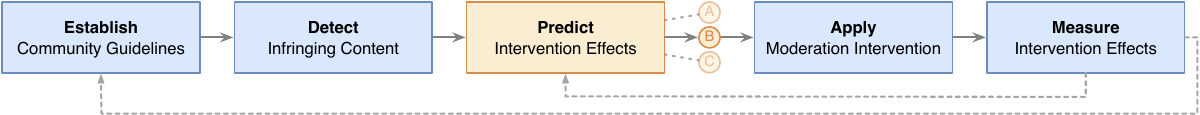}
  \caption{Overview of the moderation process with a predictive approach. Our work on predicting user abandonment following a large-scale moderation intervention is an example of how simulating different interventions and outcomes can improve platform moderation processes by opening up the possibility to take better-informed decisions.}
  \label{fig:place_in_moderation_process}
\end{figure}

Our predictive approach enhances the moderation process by enabling platform administrators to anticipate the effects of moderation interventions, allowing  more informed and effective decision-making on a tighter feedback loop, as depicted in Figure~\ref{fig:place_in_moderation_process}. In our specific case, administrators can estimate the likelihood of user abandonment after a large-scale intervention and tailor their moderation strategy accordingly. Interventions such as the collective ban of subreddits, analyzed in this study, are relatively infrequent but not uncommon. These target entire groups and communities rather than individual users or posts, usually at the behest of high-level platform administrators. Consequently, in this case the task does not have stringent real-time constraints, making it well-suited for periodic, data-driven assessments of potential outcomes of different interventions in terms of user abandonment. Moreover, the proposed approach is broadly applicable to other platforms as well, provided that historical data on past interventions is available. For instance, on Facebook, it could help predict the effects of shutting down certain groups or pages, while on Twitter/X, a similar approach could estimate the impact of banning high-profile influencers with large follower bases. Further, on the long term, such an approach could also pave the way to predictive models apt to operate in more dynamic, real-time contexts, enabling finer-grained interventions targeting individual users or posts. This shift from reactive to proactive moderation stands to enhance both the efficacy and efficiency of platform governance, fostering healthier and more sustainable digital ecosystems.

 \section{Related Works}
\label{sec:relatedworks}
The majority of existing literature on content moderation is based on descriptive analyses, while only a small subset of works have adopted a predictive approach---none of which have specifically addressed the task of estimating the effects of moderation interventions. This section summarizes and discusses both descriptive and predictive literature in content moderation, as well as some relevant works in the related area of churn prediction.

\subsection{Descriptive moderation}
\label{sec:relatedworks-desc}
In discussing the many works that took a descriptive approach to analyzing moderation interventions, we first survey studies on the Great Ban and other hard interventions, because they are the most similar to our present work. Among the works that studied the Great Ban is~\citep{trujillo2021echo}, which described the changes in activity and in-group vocabulary of those users who participated in the 15 most popular subreddits out of the 2,000 shut with the ban, using a word frequency divergence. The results highlighted the heterogeneity of the effects of the intervention and that top users tended to reduce their activity and their use of in-group vocabulary. The studies in~\citep{cima2024great,cima2025investigating} carried out similar analyses but focused on evaluating changes in toxicity rather than activity. The authors found that 15.6\% of the affected users abandoned Reddit after the ban. Among those who stayed there was a general reduction in toxic comments, but a small subset of users drastically increased their toxicity. Other works that evaluated the effects of hard content moderation include~\citep{ribeiro2024deplatforming,jhaver2021evaluating}. The former measured changes occurred in Google Trends and Wikipedia, using the Google Knowledge Graph identifiers to apply time-series analysis on the digital attention traces. The latter studied the deplatforming of Twitter influencers and its effects on the activity of their supporters.
Both works revealed that banning toxic influencers reduced the attention toward them and the number of posts citing them. These studies concluded that deplatforming may be a valid moderation strategy when used appropriately.

Due to its prominence on Reddit and the multiple interventions that it faced, \subr{The\_Donald} was the focus of many works that investigated the effectiveness of online content moderation.
For example, some works showed how the quarantine was able to reduce the visibility of the subreddit. At the same time however, it was unsuccessful at reducing racist and misogynistic language~\citep{chandrasekharan2022quarantined,shen2022tale}. Others examined the interventions in terms of activity, toxicity, factual reporting, and political bias both at community and user levels, using interrupted time series (ITS) regression analysis and Bayesian structural time series (BSTS) modeling~\citep{trujillo2022make,trujillo2023one}. 
Community-level results found a general reduction in activity, a strong long-term increase in toxicity, a slight decrease in factual reporting, and no particular change in political ideology~\citep{trujillo2022make}.
However, user-level reactions were more diversified and sometimes even extreme, especially for the most active users~\citep{trujillo2023one}. These results highlight that community-level effects are not always representative of the underlying user-level effects, which once again reaffirms the limitations of one-size-fits-all moderation~\citep{cresci2022personalized,cima2025contextualized}. On the contrary, studying, and possibly even predicting user-level reactions to moderation interventions could be particularly beneficial to moderators.

\subsection{Predictive moderation}
\label{sec:relatedworks-pred}
Recent computational efforts in the study of content moderation increasingly adopt predictive rather than descriptive approaches. Among this body of work, the vast majority of studies tackled the task of predicting which pieces of content would be subject to moderation. For example, \citep{kurdi2020video} predicted with high accuracy which YouTube videos would be later moderated using video, comment, and text-related features. The predictions were accurate even at posting-time, reducing the disappointment users face upon deletion after publication. Similarly,~\citep{paudel2023lambretta} proposed \textsc{Lambretta}, a learning to rank system developed to identify tweets likely to be removed from the platform because they convey false information. The system is intended to provide support to moderators in preventing the spread of fake news. Along the same line, \textsc{CrossMod}~\citep{chandrasekharan2019crossmod} automatically detects which Reddit posts are likely to be removed by the moderators, using cross-community learning via an ensemble of classifiers. The system is also capable of taking actions based on certain conditions set by the moderators themselves. The aforementioned works serve a twofold goal. On the one hand, they contribute to reverse engineer the content moderation actions taken by large online platforms, which often lack transparency and consistency~\citep{trujillo2023dsa}. On the other hand, the automated systems developed as part of the studies could also support human moderators by reducing the burden of extensive manual analyses as well as by reducing the emotional toll that comes from being exposed to toxic, harmful, or otherwise inappropriate content~\citep{steiger2021psychological}. Other related works focused on predicting user behaviors, such as~\citep{niverthi2022characterizing} that aimed to identify those users that would evade a Wikipedia ban by creating a new account. Additional tasks in the same work included early detection of such evasion and account matching. Instead,~\citep{habib2022proactive} discussed the feasibility of proactive moderation interventions on Reddit, looking at vocabulary features to identify problematic communities. Results showed that Reddit communities are constantly evolving, and that communities bound to become toxic can be preemptively identified. This early detection opens up the possibility to intervene before the problem escalates, thus possibly resulting in the application of less restrictive and punitive interventions.

This survey of the existing literature on predictive content moderation showed that all previous works focused on detecting which content, users, or communities would later face or evade moderation. Conversely, to the best of our knowledge, no one has focused on predicting the outcome of a moderation intervention. Our present work contributes to filling this gap.

\subsection{Churn prediction}
Churn prediction is a well-studied task in customer management that aims at forecasting whether a user will stop engaging with a platform or service based on their past behavior. It is related to our novel task of predicting user abandonment following a moderation intervention since the latter can be seen as a specific case of churn influenced by platform enforcement actions. Churn prediction has been traditionally tackled through various approaches, including the use of different feature sets, data sampling techniques, and machine learning models~\citep{geiler2022survey}. 
The majority of works tackled churn prediction in the context of telecom services. Some of them leverage Long Short-Term Memory (LSTM) and Convolutional Neural Networks (CNN)~\citep{khattak2023customer, alboukaey2020dynamic}. The former does not consider information sequence by using a Bidirectional LSTM, while the latter uses temporal features with different granularity to maximize the accuracy.
Other studies in this domain addressed data skewness through balancing techniques~\citep{sikri2024enhancing} or explored model explainability~\citep{poudel2024explaining}. Other works study user adoption and churn related to the use of web browsers~\citep{wu2022customer}, by using a new model called Multivariate Behavior Sequence Transformer (MBST) to explore temporal and behavioral information separately. Another set of works covers churn prediction for the usage of financial services~\citep{de2020incorporating,vo2021leveraging}, the former used CNNs to analyze text, while the second work used unstructured data like customer interactions as features. Although no prior research specifically addressed user retention in social media, numerous studies have examined users lifecycles within online communities. This body of work investigated various aspects of user engagement, such as how individuals interact with platforms~\citep{malinen2015understanding} or how the way their language changes over time could be an indicator of the platform evolution~\citep{danescu2013no}. Additionally, research has explored the motivations and patterns behind user migration to alternative platforms by analyzing user comments~\citep{newell2016user}. Temporal user dynamics on social media have been a key focus in studies such as~\citep{yao2021join,tardelli2024temporal}. The former traced user trajectories from their initial joining phase as newcomers, to their eventual departure from a platform. Instead, the latter investigated the evolution of coordinated communities over time during a campaign by evaluating user retention in single communities. Lastly, Cunha et al. highlight how variations in both user and community behaviors contribute to different levels of community success~\citep{cunha2019all}.
 \section{Problem Definition}
\label{sec:problem}
We introduce the new task of predicting whether a moderation intervention will lead a user to abandon the platform. We first outline the general framing of the problem before specifying our classification task in the context of Reddit’s Great Ban.

\begin{figure}[t]
  \centering
  \includegraphics[width = 0.7\linewidth]{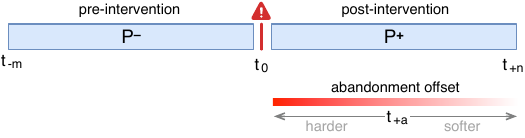}
  \caption{Problem definition time frame. Given an intervention at $t_0$, we want to predict user abandonment in $P^+$, based on user data from $P^-$ and inactivity during a time window from offset $t_{+a}$ to $t_{+n}$.}
  \label{fig:problem_time_frame}
\end{figure}

\subsection{Problem framing}
\label{sec:problem-framing}
Let $i$ denote a moderation intervention that takes place at time $t_0$. To analyze user behavior around this event, we define three key time periods, as shown in Figure~\ref{fig:problem_time_frame}:
\begin{itemize}
    \item The \textbf{pre-intervention period} $P^- = [t_{-m}, t_0)$ represents user activity before the intervention, where $t_{-m} < t_0$.
    \item The \textbf{post-intervention period} $P^+ = (t_0, t_{+n}]$ captures user activity after the intervention, where $t_0 < t_{+n}$.
    \item The \textbf{abandonment period} $P^a = [t_{+a}, t_{+n}]$, with $t_0 < t_{+a} < t_{+n}$, is a subset of $P^+$ used to determine whether a user has disengaged from the platform.
\end{itemize}
Note that neither $P^-$ nor $P^+$ include $t_0$, since user activity at the exact moment of the intervention may be non-representative due to sudden fluctuations or spurious activity spikes. For any given user $u$, we define their activity during the aforementioned periods as follows:
\begin{itemize}
    \item $A_u^-$ is the \textbf{pre-intervention activity} of user $u$ during $P^-$.
    \item $A_u^+$ is the \textbf{post-intervention activity} of user $u$ during $P^+$.
    \item $A_u^a$ is the activity of user $u$ during the abandonment period $P^a$.
\end{itemize}
Using these definitions, we frame the task of predicting user abandonment as a binary classification problem, with the following two classes:
\begin{align*}
    C_{+1} &= \{u \; | \; A_u^- > 0 \; \text{and} \; A_u^a = 0\} \quad \text{(\texttt{abandoning} users)} \\
    C_{-1} &= \{u \; | \; A_u^- > 0 \; \text{and} \; A_u^a > 0\} \quad \text{(\texttt{non-abandoning} users)}.
\end{align*}
The positive class $C_{+1}$ represents users who were active before the intervention but exhibited no activity during $P^a$, indicating abandonment. The negative class $C_{-1}$ consists of users who remained active during both $P^-$ and $P^a$, indicating continued engagement with the platform. The goal of our predictive task is to learn a function $f$ that assigns the correct class $C_u$ to each user $u$ based on their pre-intervention activity $A_u^-$:
\begin{equation}
    \label{eq:hard-abandonment}
    f(A_u^-) \approx C_u \in \{C_{+1}, C_{-1}\}.
\end{equation}
This formulation allows assessing whether pre-intervention behavior provides sufficient information to anticipate user abandonment following a moderation intervention.

Having defined the general framing of the task, we now introduce the distinction between different types of abandonment. In reality, user activity exists on a spectrum and there is no absolute boundary between full abandonment and continued participation. Some users may abruptly disengage after an intervention, others may do it gradually, while others still may reduce participation dramatically but nonetheless post sporadically. Whether certain users should be classified as abandoning or not revolves around defining a threshold for inactivity. The abandonment offset $t_{+a}$ addresses this by setting a temporal boundary: if a user exhibits no activity beyond $t_{+a}$, they are considered to have abandoned the platform. For example, this choice helps discount any brief, reactionary activity that may occur immediately after the intervention. Therefore, a shorter $t_{+a}$ captures \textit{short-term abandonment}, where users disengage almost immediately following the intervention. Conversely, a longer $t_{+a}$ accounts for \textit{longer-term abandonment}, recognizing users who may have remained somewhat active for a short period before eventually leaving the platform. Selecting an appropriate $t_{+a}$ provides flexibility in defining abandonment, making it possible to tailor the classification task to different analytical and practical needs.

\subsection{Task specification}
\label{sec:problem-tasks}
In this work, we aim to predict those users of banned subreddits who abandoned the platform ---that is, who ceased activity on Reddit--- after the Great Ban. Given the novelty of the problem and for the sake of simplicity, we perform two variants of the classification task $C$. In both variants we set $m=n$ so that $P^-$ and $P^+$ have the same duration, but we use two different settings for $t_a$, which results in two different abandonment periods $P^a$: 
\begin{align*}
    C^{H} &:= \{C \; | \; t_{+a} = t_0 \; \implies \; P^a=P^+\} \quad \text{(\texttt{hard abandonment})} \\
    C^{S} &:= \{C \; | \; t_{+a} \gg t_0 \; \implies \; P^a \subset P^+\} \quad \text{(\texttt{soft abandonment})}.
\end{align*}
The hard abandonment task $C^{H}$ follows a strict definition, where a user is considered non-abandoning if they exhibit \textit{any} activity on the platform after the intervention. Conversely, the soft abandonment task $C^{S}$ introduces a more flexible criterion. In this case, a user may continue posting for a limited time after the intervention ---up to $t_{+a}$--- and still be classified as abandoning if they become inactive beyond this point. This approach allows for a more nuanced distinction, recognizing that some users may gradually disengage rather than leaving the platform immediately.

Based on previous research that analyzed the medium-term effects of moderation interventions on Reddit~\citep{trujillo2022make,cima2024great}, we determined the duration of $P^-$ and $P^+$ to be 210 days, for a total time frame of 421 days (including the day $t_0$ of the Great Ban). The selected time frame provides a convenient and practical period that can be divided into months (7 months) or weeks (30 weeks). Hence, for $C^H$ the duration of $P^a=P^+=$ 7 months. On the other hand, for $C^S$ we selected the offset $t_{+a}=$ 4 months, with the duration of the corresponding abandonment period $P^a=$ 3 months (92 days), based on a precursory analysis of different offsets and class balances between non-abandoning and abandoning users. The 4-month inactivity threshold balances a long enough time to indicate meaningful disengagement while avoiding possible short-term fluctuations in user activity.
 \begin{table}[t]
\small
\centering
\setlength{\tabcolsep}{3pt}
\begin{tabular}{ l r r r }
\toprule
 
\textbf{subreddit} & \textbf{subscribers} & \textbf{users} & \textbf{comments}\\
\midrule
\subr{ChapoTrapHouse} & 159,185 & 9,295 & 1,368,874 \\
\subr{The\_Donald} & 792,050 & 4,262 & 619,434 \\
\subr{DarkHumorAndMemes} & 421,506 & 1,632 & 35,561 \\ 
\subr{ConsumeProduct} & 64,937 & 1,730 & 60,073 \\
\subr{GenderCritical} & 64,772 & 1,091 & 94,735 \\
\subr{TheNewRight} & 41,230 & 729 & 5,792 \\
\subr{soyboys} & 17,578 & 596 & 5,102 \\
\subr{ShitNeoconsSay} & 8,701 & 559 & 9,178 \\
\subr{DebateAltRight} & 7,381 & 488 & 27,814 \\
\subr{DarkJokeCentral} & 185,399 &316 &  3,214\\
\subr{Wojak} &  26,816 & 244 & 1,666\\ 
\subr{HateCrimeHoaxes} &  20,111 & 189 & 775\\
\subr{CCJ2} & 11,834 & 150 & 9,785\\
\subr{imgoingtohellforthis2} & 47,363  & 93 &376 \\
\subr{OandAExclusiveForum} & 2,389 & 60 & 1,313 \\ 
\bottomrule
\end{tabular}
\caption{List of the 15 banned subreddits used for the analysis, sorted by number of active users. Data from these subreddits constitutes dataset \textbf{D}\textsubscript{B-B}. 
}
\label{tab:dataset}
\end{table}

\section{Data Collection and Preparation}
\label{sec:dataset}
For this study we use the dataset from~\citep{cima2024great,cima2025investigating}, consisting of 16 million Reddit comments made by 16,828 different users that were affected by the Great Ban.

\subsection{Banned subreddits selection}
Each user in the dataset was an active participant before the ban in at least one of 15 selected subreddits. The selection of the 15 subreddits was done as follows. Out of the 2,000 subreddits shut during the Great Ban, Reddit only publicly disclosed the names of the 10 largest ones.\footnote{\url{https://www.redditstatic.com/banned-subreddits-june-2020.txt} (accessed: 03/15/2024)} For all remaining banned subreddits, Reddit only provided a list of partially obfuscated names. M. Trujillo et al.~\citep{trujillo2021echo} deciphered the list, removed all private subreddits as well as all those having less than 2,000 active users. Only 5 subreddits remained from the obfuscated list, which M. Trujillo et al.~\citep{trujillo2021echo} and Cima et al.~\citep{cima2024great} studied in addition to the 10 publicly disclosed ones. Table~\ref{tab:dataset} provides the list of the 15 subreddits on which our study is focused, together with some descriptive statistics. 
\begin{figure}[t]
  \includegraphics[width = 0.6\linewidth]{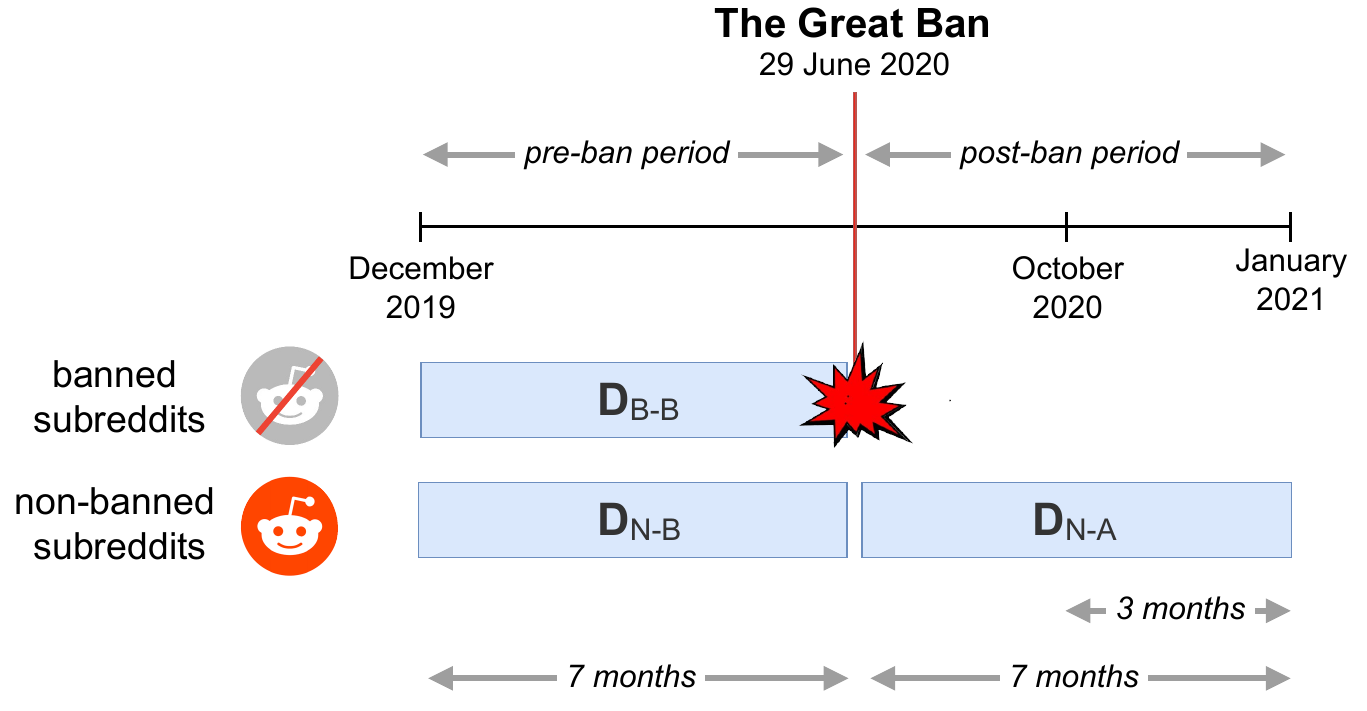}
  \centering
  \caption{Composition of our datasets and data collection periods. \textbf{D}\textsubscript{B-B}: data from within the banned subreddits, before the ban. Dataset \textbf{D}\textsubscript{B-B} is used to select representative users from the banned subreddits. \textbf{D}\textsubscript{N-B}: data from non-banned subreddits, before the ban. Datasets \textbf{D}\textsubscript{B-B} and \textbf{D}\textsubscript{N-B} (i.e., both pre-ban datasets) are used to compute machine learning features. \textbf{D}\textsubscript{N-A}: data from non-banned subreddits, after the ban. Dataset \textbf{D}\textsubscript{N-A} is used to provide ground-truth labels for the users based on their activity post-ban.}

  \label{fig:groundtruth_timeline}
\end{figure}

\subsection{Active users selection}
For each of the 15 selected subreddits, Cima et al.~\citep{cima2024great} gathered all comments posted therein between December 2019 and June 2020, which resulted in a total of 8 million comments posted by approximately 194K distinct users. As shown in Figure~\ref{fig:groundtruth_timeline}, the timeframe used for the data collection spans 7 months leading up to the Great Ban, providing a strong reference for the activity levels of the affected users before the moderation intervention~\citep{trujillo2023one}. The initial set of 194K users was later filtered so as to retain only those users who showed consistent activity in at least one of the 15 selected subreddits. In detail, only those users who posted at least one comment each month were retained. In~\citep{cima2024great}, this filtering step was useful to obtain meaningful post-hoc estimations of the effect of the Great Ban on the activity of the users. Since here we also study changes in user activity, albeit in a predictive fashion, this filtering step is suitable for our analysis as well. In addition, bots were also removed from the dataset. First, we followed reference practices in literature and we removed all accounts that posted two or more comments in less than a second~\citep{hurtado2019bot}. Subsequently, we manually analyzed a random sample of 1,000 removed accounts to ensure that only bots were removed from the dataset. A similar analysis also revealed that among the remaining users in the dataset there were no clearly distinguishable bots. As a result of these filtering and validation steps, we obtained the dataset \textbf{D}\textsubscript{B-B} composed of 2.2 million comments posted within the 15 selected subreddits by 16,828 distinct users, as summarized in Table~\ref{tab:dataset}.

\subsection{Data from outside the banned subreddits}
Evaluating (or predicting) the effect of a moderation intervention involves comparing data from before and after the intervention itself. This requirement also surfaces from the definitions that we gave in Section~\ref{sec:problem}. However, no activity exists post-intervention within the banned subreddits, since they were all permanently shut. Therefore, data about the behavior of the affected users \textit{outside} of the banned subreddits must be used. For this reason, Cima et al.~\citep{cima2024great} collected all comments made by the 16,828 selected users outside of the 15 banned subreddits over a wide timeframe spanning 7 months before and 7 months after the Great Ban, as depicted in Figure~\ref{fig:groundtruth_timeline}. This additional data collection yielded approximately 13.8 million comments, of which 8.2 million (59\%) were posted before the ban and constitute dataset \textbf{D}\textsubscript{N-B} while 5.6 million (41\%) were posted afterwards and constitute dataset \textbf{D}\textsubscript{N-A}. We used the data from before the ban (\textbf{D}\textsubscript{N-B}) to compute our machine learning features, while the data posted after the ban (\textbf{D}\textsubscript{N-A}) allows assigning ground-truth labels to the users based on their post-ban activity. Notably, out of the 16,828 initially selected users, 288 (1.7\%) did not have any activity outside of the 15 banned subreddits and were thus discarded.  \begin{figure*}[t]
  \includegraphics[width=\textwidth,keepaspectratio]{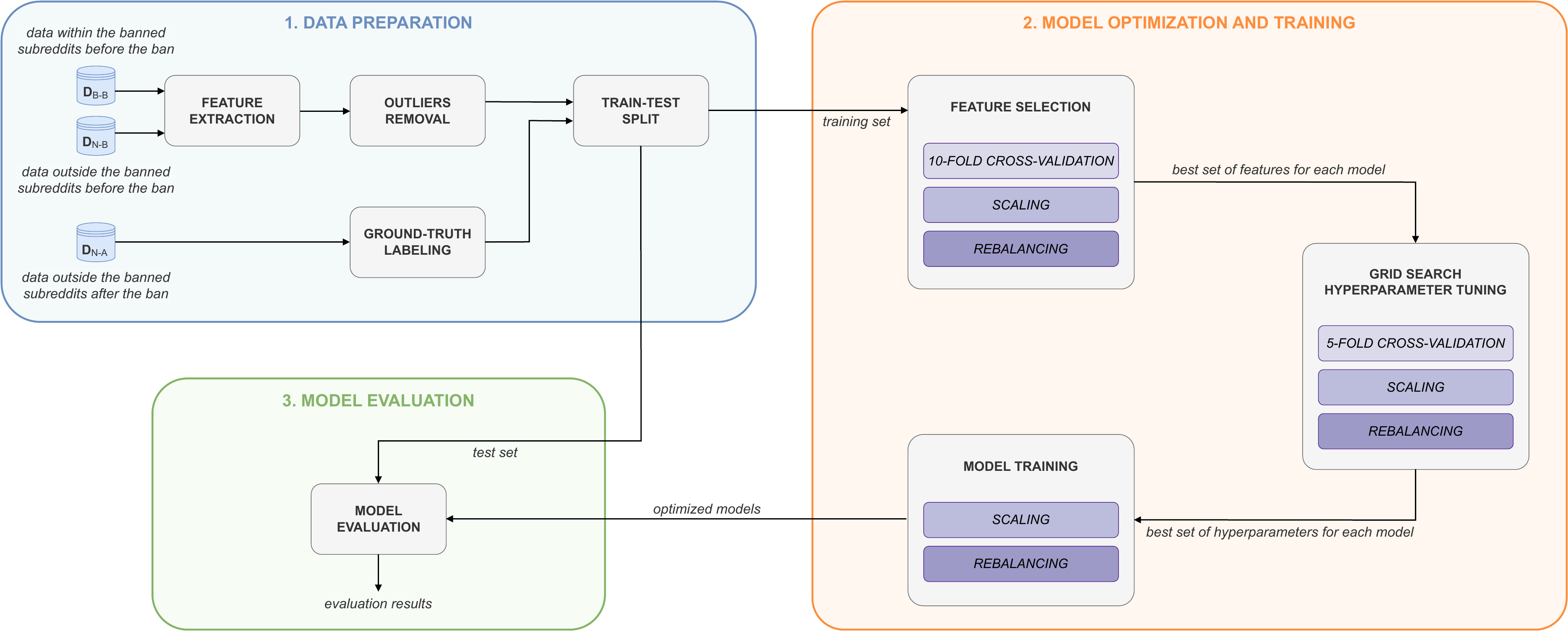}
  \caption{Machine learning pipeline. Our data preparation steps involve feature extraction, outliers removal, ground-truth labeling, and splitting of the dataset into a training and a test set. The training set is used for model training and optimization. This involves feature scaling and selection, data rebalancing, and hyperparameters optimization. Finally, the optimized models are evaluated on the held-out test set.}
  \label{fig:ml}
\end{figure*}

\section{Machine Learning Approach}
\label{sec:approach}
The formulation of hard and soft abandonment prediction given in Section~\ref{sec:problem-tasks} cast both problems as binary classification tasks. Our approach to these tasks is sketched in Figure~\ref{fig:ml} and summarized in the remainder of this section, while the implementation details and the experimental settings are described in Section~\ref{sec:experiments}.

We begin by extracting user-level features from the portion of the dataset related to the activity of the selected users before the ban, both within (\textbf{D}\textsubscript{B-B}) and outside (\textbf{D}\textsubscript{N-B}) of the banned subreddits. Before further processing this data we perform outliers removal, as this helps enhancing the robustness and generalization of the trained models by mitigating the potential distortions and biases caused by extreme data points~\citep{smiti2020critical}. Data about user activities outside of the banned subreddits after the ban (\textbf{D}\textsubscript{N-A}) is used to assign ground-truth labels to the users. 
User-level features are then merged with the ground-truth labels and the resulting dataset is split into a training and a test set. The test set is only used for the final evaluation of the optimized models. Instead, the training set is used for feature selection and for optimizing and training the models. Given that we experiment with multiple classification models, each with its own characteristics and hyperparameters, the feature selection and optimization steps of our machine learning pipeline ensure that each model operates in optimal conditions. In other words, our approach ensures meaningful and fair comparisons between the different models. The first step in the model optimization process involves selecting an adequate number and set of features for each model. Specifically, we rank all features and we select the top-\textit{N} ones to reduce the dimensionality of our models and to eliminate redundant and correlated features. For each model we experiment with multiple numbers \textit{N} of selected features, as different models might benefit from a smaller or larger number of features. To have accurate and robust estimates of the model performances with the different sets of features, we carry out a 10-fold cross-validation on the training set. At each iteration, we rescale features so that they all cover a similar range of values. Rescaling ensures that no feature dominates the learning process due to overly large magnitude, and also contributes to speeding up training times ~\citep{ozsahin2022impact,wan2019influence}. At each iteration we also rebalance the training set. In fact, both the hard and soft abandonment tasks are heavily imbalanced, given that abandoning users constituted a small minority of all users (i.e., 14.9\% for hard abandonment and 26.9\% for soft abandonment). We note that this imbalance is common for tasks concerning the prediction of human behavior~\citep{robertson2022uncommon}. Therefore, by rebalancing the training set we reduce the bias that the trained models might exhibit towards those users who did not abandon the platform (i.e., the majority class). The averaged results of the 10-fold cross-validation process are used to select the optimal number and set of features for each model. 
Up to now, all models were trained and evaluated with default hyperparameters. In the last step of model optimization, for each model we perform a grid search with cross-validation over its hyperparameters so as to choose the best combination of them.  Finally, we use the selected set of features and hyperparameters to train an optimized version of each model on the whole training set. Feature scaling and rebalancing are applied also for hyperparameter tuning and model training. Each optimized model is then evaluated on the held-out test set. Evaluation results and comparisons are presented in Section~\ref{sec:results}.
 \section{Feature Engineering}
\label{sec:features}
We compute a total of 142 features for each user in the dataset. Our features are organized into the following four main classes: 30 (21\%) \textit{activity} features; 40 (28\%) \textit{toxicity} features; 36 (25.5\%) \textit{writing style} features; and 36 (25.5\%) \textit{relational} features.

\subsection{Activity features}
Features in this class describe the overall level of engagement and participation of a user within Reddit. 
The motivation for including activity features is straightforward, given our goal of predicting user abandonment---that is, the interruption of posting activity on the platform. Moreover, previous descriptive studies on the effects of moderation interventions, such as those reviewed in Section~\ref{sec:relatedworks-desc}, showed that bans frequently cause a reduction in user activity~\citep{cima2024great,trujillo2021echo,trujillo2022make,cima2025investigating}, which supports the adoption of this type of features in our task. Some examples of the activity features that we implemented are the total number of comments posted by a user, the average time between the sharing of two subsequent comments, the slope of the trend of posted comments in either the banned and the non-banned subreddits, which captures whether user activity in certain subreddits was increasing or decreasing prior to the ban. Other features of this class capture possible peculiar characteristics of the comments, one of them being the number of \textit{stickied} comments. This feature was included as it was among the strongest predictors of ban evasion in a previous study~\citep{habib2022proactive}. 

\subsection{Toxicity features}
The users considered in our study were active participants in subreddits banned due to the widespread presence of toxic content. As such, features gauging the degree of toxicity of the users pre-ban could be strong predictors in our task. In addition, integrating toxicity features into the classification models allows exploring the extent to which differences in toxic behavior are capable of explaining user reactions to the ban. We computed toxicity scores with \textsc{Detoxify}~\citep{Detoxify}, a state-of-the-art~\citep{cima2024great} multilingual deep learning toxicity classifier that is widely used in literature~\citep{bagdasaryan2022spinning,kopf2024openassistant}, including predictive studies for the detection of cyberbullying~\citep{teng2023cyberbullying,ejaz2024towards}. Given a piece of text, \textsc{Detoxify} provides multiple indicators of toxicity, such as the \textit{toxicity} and \textit{severe toxicity} scores, as well as additional scores for \textit{obscenity}, \textit{insults}, \textit{identity attacks}, and \textit{threats}. In addition, we also computed sentiment scores given the strong relationship between toxicity and sentiment~\citep{gilda2022predicting}. For obtaining sentiment scores we used \textsc{Vader}, a well-known rule-based sentiment analyzer that is specifically designed for social media content~\citep{vader}. \textsc{Vader} outputs \textit{positive}, \textit{negative}, \textit{neutral}, and \textit{compound} sentiment scores, with the latter being the sum of the previous ones. To compute toxicity and sentiment user-level features we first classified each comment by each user with \textsc{Detoxify} and \textsc{Vader}. We then aggregated each score provided by the two tools over all comments by the same user, by computing the \textit{mean}, \textit{minimum}, \textit{maximum}, and \textit{standard deviation} of the scores, obtaining the $4 \times 10 = 40$ features in this class.

\subsection{Writing style features}
Several previous works demonstrated that users may exhibit linguistic changes after suffering a moderation intervention. For example,~\citep{horta2021platform} noted changes in the use of the first and third plural pronouns, as well as in word choices, after a ban. Similarly,~\citep{trujillo2021echo} found that users who remained on Reddit after a ban exhibited reduced use of in-group language. Other prior research has shown that moderation interventions can lead to linguistic shifts, including changes in word choices, sentiment, and rhetorical strategies. Studies have observed that moderated users may adjust their language to evade detection, adopt more aggressive or explicit speech in response to relaxed moderation policies, or even develop lexical variations of restricted terms to bypass enforcement mechanisms. These findings suggest that subtle linguistic cues may reflect user adaptation, disengagement, or defiance following an intervention~\citep{chancellor2016thyghgapp,trujillo2021echo,wadden2021effect}. Based on these preliminary descriptive results, here we assess the extent to which basic linguistic and writing style features provide predictive information about the activity of the users after a ban. Among the features that we computed are counts of the different \textit{parts-of-speech} used and readability scores such as the well-known \textit{Flesch-Kincaid Grade Level} and the \textit{SMOG Index}. Each of these scores were computed for each user comment and subsequently aggregated at the user level by computing the \textit{mean}, \textit{minimum}, \textit{maximum}, and \textit{standard deviation}.

\subsection{Relational features}
The relationships between users and communities constitute the fundamental fabric upon which social media platforms are woven, shaping the dynamics of online engagement and interaction. Moreover, changes in user-user and user-subreddit relationships have already been observed as a consequence of moderation interventions~\citep{trujillo2022make,habib2022proactive}. For this reason, considering relational features in our task allows us to capture the interaction dynamics that are inherent to Reddit communities. As an example, we computed features that quantify the degree of influence that each user had in the banned subreddits, which could be predictive of future abandonment. This feature was computed by ranking users based on the average score (i.e., the difference between upvotes and downvotes) of their comments in the banned subreddits. In addition, the same feature was also computed for the non-banned subreddits. Similarly, we examined the relationship between users and specific subreddits, identifying those users who predominantly participated in the banned subreddits, which could provide an early sign of abandonment. In fact, users who almost exclusively participated in banned subreddits could lack motivation to stay on the platform after those subreddits were shut~\citep{horta2021platform}. Finally, we leveraged results of recent works that demonstrated the usefulness of \textit{initiative} and \textit{adaptability} features in social media user classification tasks~\citep{mazza2022investigating}. For example, we measured the number and ratio of threads started by each user as a proxy for their capacity to drive the conversation in a subreddit, rather than to follow what others say. 
 \section{Experiments and Settings}
\label{sec:experiments}

\subsection{Comparisons}
\label{sec:experiments-comparisons}
We experiment with multiple reference classification models to solve the hard and soft abandonment tasks, including Naive Bayes (\texttt{NB}), K-Nearest Neighbors (\texttt{KNN}), Decision Tree (\texttt{DT}), Random Forest (\texttt{RF}), Adaptive Boosting (\texttt{AB}), Gradient Boosting (\texttt{GB}), and Support Vector Machine (\texttt{SVM}). This selection spans the main families of classification algorithms, encompassing instance-based, probabilistic, ensemble-based, and discriminative methods of different complexities, thus providing an extensive exploration of many learning paradigms. Furthermore, similar choices of classification models were made in recent works on related tasks, such as the detection of different forms of online harms~\citep{cresci2015fame,liu2016detecting,tardelli2022detecting,amira2023detection}. In addition to the previous classification models, we also implement five simple baselines for further comparison. The \texttt{Stratified} baseline generates predictions by following the ground-truth label distribution, without taking any feature into account. The \texttt{DT Ratio} baseline employs a decision tree that receives a single feature in input representing the ratio of comments made by a user in the non-banned subreddits over the comments made by that user in the banned ones. The rationale for this baseline is that users who are more active on non-banned subreddits are less prone to abandon the platform. The \texttt{DT Trend} baseline employs a decision tree that receives a single feature in input representing the trend of the number of monthly comments by a user before the ban. The reason for including this baseline is that users with a decreasing posting trend are more likely to abandon the platform. Finally, to test the effectiveness of our machine learning pipeline, we train a Naive Bayes (\texttt{NB-NFE}) and a Decision Tree (\texttt{DT-NFE}) model by only performing feature standardization, without any other step of our pipeline ---namely, rebalancing, feature selection, and hyperparameter tuning. For the sake of simplicity, we do not apply any preprocessing steps to the baselines. Finally, we remark that we resort to using these baselines instead of more powerful approaches because the novelty of our task makes it so that no previous work exists on predicting the effects of a moderation intervention.
\subsection{Machine learning pipeline}
In the following we report the experimental settings and implementation details of our machine learning pipeline, showed in Figure~\ref{fig:ml}. 
\subsubsection{Outliers removal}
To carry out outliers detection we apply \textit{Isolation Forest}, a decision tree-based method that is particularly suitable for identifying anomalies in highly-dimensional data, such as ours~\citep{liu2008isolation}. Isolation Forest constructs decision trees by randomly selecting features and split points, efficiently isolating anomalies by assigning them shorter average path lengths in the trees compared to normal data instances, by leveraging the natural tendency of anomalies to require fewer splits for isolation. The application of Isolation Forest to our dataset led to the detection and removal of 297 outliers. A manual verification revealed that the removed users featured extremely low levels of activity, which supports their removal. 
\subsubsection{Train-test split}
We use a common 80/20 split of our dataset into the training and test sets, following the Pareto principle. Data in the two sets is stratified according to the class labels.
\subsubsection{Cross-validation}
During model optimization we perform two stratified cross-validations on the training set to obtain robust estimates of model performance. In detail, we perform a 10-fold cross-validation during feature selection and a 5-fold cross-validation during hyperparameter tuning. The stratification is necessary given our marked class imbalance, so that each class is represented proportionally in both the training and validation sets of the cross-validation.

\subsubsection{Feature scaling and rebalancing}
\label{sec:exp-scaling-rebalancing}
During the feature selection, hyperparameter tuning, and model training steps, we perform feature scaling for all models. Additionally, for all models except for the ensemble-based ones ---namely, \texttt{RF}, \texttt{AB}, and \texttt{GB}--- we also perform data rebalancing. For ensemble-based models, removing data rebalancing often improves performance, as also confirmed by our experimental results, because these models naturally handle class imbalance and can be negatively affected by data rebalancing, which might distort data distributions~\citep{khan2024review}. Feature scaling is applied first, via \textit{Z-score standardization}, so that each feature has mean $\mu = 0$ and standard deviation $\sigma = 1$. Rebalancing is applied last, as suggested in previous literature~\citep{smote}. For rebalancing, we leverage both oversampling and undersampling techniques to mitigate the large class imbalance of our dataset. We perform oversampling with \textit{SMOTE}, a state-of-the-art technique that creates new artificial samples of the minority class~\citep{smote}. Instead, we use \textit{Random Undersampling} to randomly discard samples from the majority class~\citep{hasanin2018effects}. For both tasks we use the two techniques in combination to reach a final fixed imbalance of 60/40 between the majority and minority class, down from 85/15 for hard abandonment and 73/27 for soft abandonment. This approach allows us to reduce the skewness of the class distribution without creating too many artificial samples that could introduce bias, and without discarding too much data that could result in degraded model performance. 
\subsubsection{Feature selection}
We perform feature selection for all models except for the ensemble-based ones. Indeed, the latter perform internal feature selection, reducing the need for explicit preprocessing, and are robust to redundant features, unlike simpler models~\citep{bolon2019ensembles}. Feature selection is achieved with \textit{ANOVA F-value}, a technique that was shown to achieve good performance in recent works on the detection of malicious online user activities~\citep{ilias2021detecting}. The ANOVA tests are used to obtain an F-value for each feature, expressing the likelihood for the feature to be predictive for the task at hand. Then, features are ranked based on their respective F-value and the top-\textit{k} are selected. Out of 142 total features, for each trained model we vary \textit{k} from 10 to 80 with a step of 10, and evaluate the performance of the resulting models as part of the 10-fold cross-validation. At the end of the 10-fold cross-validation process, for each model we select the number and set of features that maximizes its F1 score on the positive class, so that for each model, only its best set of features is used in the following steps.
\subsubsection{Hyperparameter tuning}
For each model, we perform hyperparameter tuning with a grid search 5-fold cross-validation on the training set. 
\subsubsection{Model training}
Finally, we train each model on the whole training set by using the best set of features and hyperparameters that we determined.

\subsection{Model evaluation}
We evaluate the optimized models and the baselines on the test set by means of standard evaluation metrics for classification tasks. Specifically, we report the area under the receiver operating characteristic curve (\textit{AUC}), the overall \textit{micro F1 score}, as well as the \textit{precision}, \textit{recall}, and \textit{F1 score} of the positive class (i.e, abandoning users), given its importance for moderators and platform administrators in the context of planning moderation interventions. We also report the 95\% confidence intervals of the \textit{positive} \textit{F1} and the overall \textit{micro F1} scores.
 \begin{table}[t]
    \small
    \centering
    \setlength{\tabcolsep}{2.5pt}
    \begin{tabular}{clcccccccc}
        \toprule
        \multicolumn{3}{c}{} & \multicolumn{4}{c}{\textbf{positive class}} & \multicolumn{3}{c}{\textbf{overall}} \\ 
        \cmidrule(r){4-7}\cmidrule(l){8-10}
        \textbf{task} & \textbf{model} & \textit{p} & \textit{precision} & \textit{recall} & \textit{F1} & \textit{95\% CI} &  \textit{AUC}  & \textit{micro F1} & \textit{95\% CI} \\
        \midrule
            \multirow{13}{*}{\rotatebox[origin=c]{90}{\textbf{hard abandonment}}} & \textit{baselines} \\ [0.4em]
            & \texttt{Stratified} & -- & 0.155 & 0.158 & 0.157 & - & 0.501 & 0.751 & -\\ 
            & \texttt{DT Ratio} & 1  & 0.356 & 0.199 & 0.255 & - & 0.537 & 0.689 & - \\
            & \texttt{DT Trend} & 1 &  0.325 & 0.219 & 0.262 & - & 0.591 & 0.726 & -\\
             & \texttt{NB-NFE} & 142 & 0.206 & 0.313 & 0.248 & - & 0.595  & 0.814 & - \\
             & \texttt{DT-NFE} & 142 & \underline{0.636} & 0.230 & 0.338 & - & 0.630  & 0.627 & - \\
            \cmidrule(r){2-10}
            & \textit{trained models} \\ [0.4em]
            & \texttt{KNN} & 20 & 0.514 & 0.308 & 0.384 & [0.332, 0.363] & 0.719 & 0.754 & [0.703, 0.715] \\
            & \texttt{NB}  & 10 & 0.294 & 0.300 & 0.297 & [0.309, 0.345] & 0.680 & 0.792 & [0.795, 0.812] \\
            & \texttt{RF}  &142 & 0.634  & \underline{0.723}  & \underline{0.675} & [0.380, 0.600]  & 0.909  & \underline{0.909} & [0.862, 0.895] \\
            & \texttt{AB}  &142 & 0.630  & 0.691  & 0.659 & [0.412, 0.597] & \underline{0.918}  & 0.902 & [0.768, 0.890] \\
            & \texttt{DT} & 10 & 0.617 & 0.330 & 0.430 &  [0.315, 0.360] & 0.752 & 0.756 & [0.710, 0.740] \\
            & \texttt{GB}  &142 & \textbf{0.658}  & \textbf{0.736}  & \textbf{0.695}  & [0.397, 0.608] & \textbf{0.930}  & \textbf{0.914} & [0.719, 0.891] \\
            & \texttt{SVM} & 50 & 0.520 & 0.373 & 0.434  & [0.437, 0.465] & 0.781 & 0.797 & [0.803, 0.817]\\
        \midrule
            \multirow{13}{*}{\rotatebox[origin=c]{90}{\textbf{soft abandonment}}} & \textit{baselines} \\ [0.4em]
            & \texttt{Stratified} & -- & 0.263 & 0.266 & 0.264 & - & 0.509 & 0.605 & -\\
            & \texttt{DT Ratio} & 1 & 0.374 & 0.328 & 0.349 & - & 0.527 & 0.624 & - \\
            & \texttt{DT Trend} & 1 & 0.282 & 0.319 & 0.299 & - & 0.549 & 0.644 & -\\
            &  \texttt{NB-NFE} & 142 & 0.091 & 0.540 & 0.156 & - & 0.604 & 0.734 & - \\
            & \texttt{DT-NFE} & 142 & \textbf{0.589} & 0.287 & 0.386 & - & 0.524 & 0.494 & -\\
            \cmidrule(r){2-10}
            & \textit{trained models} \\ [0.4em]
            & \texttt{KNN} & 10 & 0.513 & 0.438 & 0.473 & [0.428, 0.462] & 0.698 & 0.691 & [0.644, 0.664] \\
            & \texttt{NB}  & 10 & 0.400 & 0.485 & 0.438 &  [0.413, 0.450] & 0.685 & 0.723 & [0.778, 0.787] \\
            & \texttt{RF}  &142 & 0.473  & \textbf{0.744}  & 0.578 & [0.522, 0.572] & 0.801  & \textbf{0.814}& [0.745, 0.800] \\
            & \texttt{AB} &142 & 0.504  & \underline{0.706}  & \underline{0.588}  & [0.525, 0.574] & \underline{0.806}  & \underline{0.809} & [0.722, 0.815] \\
            & \texttt{DT}  & 20 & \underline{0.558} & 0.443 & 0.494 & [0.387, 0.423] & 0.695 & 0.690 & [0.610, 0.629] \\
            & \texttt{GB}  &142 & 0.513  & 0.694 & \textbf{0.590} & [0.517, 0.579]  & \textbf{0.816} & 0.808 & [0.639, 0.788] \\
            & \texttt{SVM} & 50 & 0.484 & 0.513 & 0.499 & [0.490, 0.515] & 0.728 & 0.737 & [0.736, 0.751]\\
        \bottomrule
    \end{tabular}
    \caption{Classification results for the hard abandonment (top rows) and soft abandonment (bottom rows) tasks. Column \textit{p} reports the number of features used by each model. Model performance is measured in terms of \textit{precision}, \textit{recall} and \textit{F1} on the positive class, and in terms of overall \textit{AUC} and \textit{micro F1}. For \textit{positive F1} and overall \textit{micro F1} we also report the 95\% confidence intervals based on the scores obtained in the k-fold cross-validation.} For each task, the best result in each evaluation metric is shown in \textbf{bold} and the second-best is \underline{underlined}.
    \label{tab:results-classification}
\end{table}
 
\begin{figure*}[t]
    \centering
    \begin{subfigure}[t]{0.35\columnwidth}\centering
        \includegraphics[width=\columnwidth]{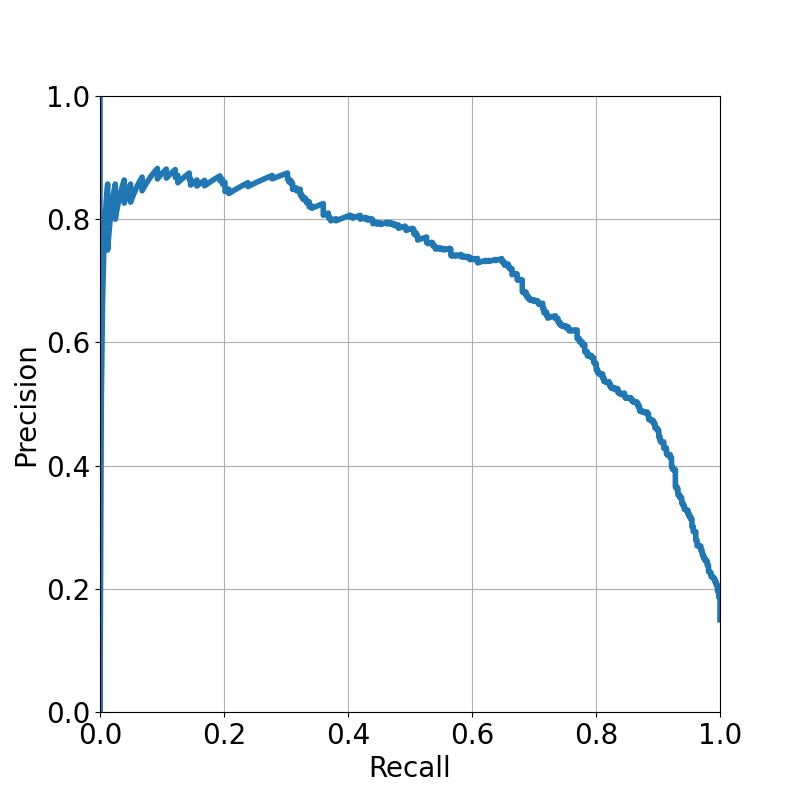}
        \caption{Hard abandonment.}
        \label{fig:pr_hard}\end{subfigure}
    \hspace{0.1\columnwidth}
    \begin{subfigure}[t]{0.35\columnwidth}\centering
        \includegraphics[width=\columnwidth]{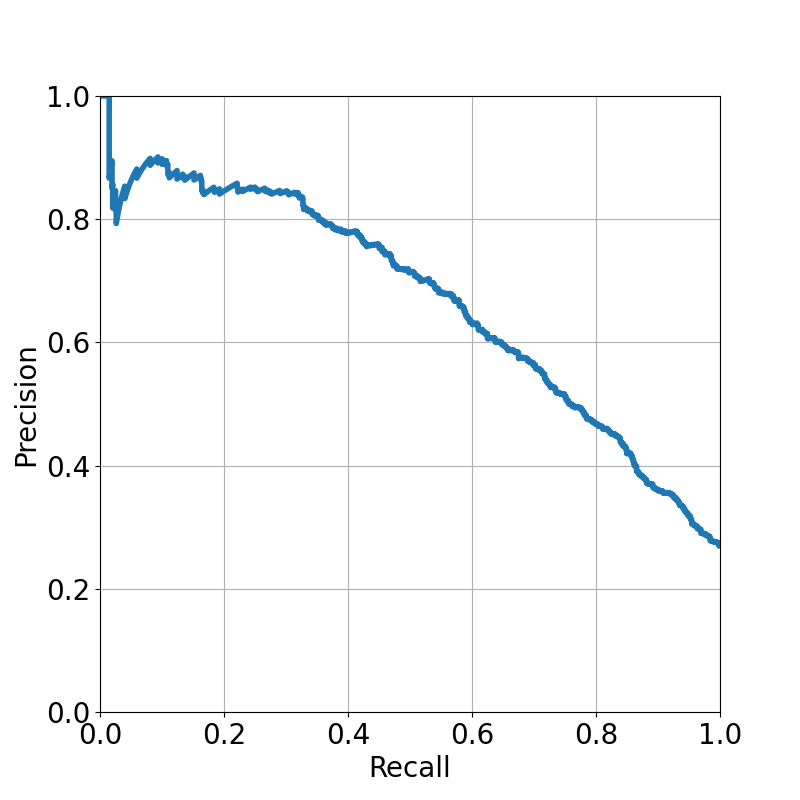}
        \caption{Soft abandonment.}
        \label{fig:pr_soft}\end{subfigure}
    \caption{Precision-Recall curves for the Gradient Boosting (\texttt{GB}) model in predicting hard and soft abandonment. The curves illustrate the trade-off between precision and recall, allowing for an assessment of how well the model balances false positives and false negatives in both tasks.}
    \label{fig:pr_curves}
\end{figure*}

\section{Results}
\label{sec:results}
\subsection{Prediction of hard and soft abandonment}
Classification results for the hard abandonment task are reported in the top half of Table~\ref{tab:results-classification}, while those for the soft abandonment task are reported in the bottom half of the table. For each task we report the performance of the baselines and of the optimized version of each trained model. For each baseline and model, Table~\ref{tab:results-classification} shows the number of used features (\textit{p}) and the evaluation metrics computed for the positive class and for both classes (\textbf{overall} columns).
With respect to the hard abandonment task, Gradient Boosting (\texttt{GB}) achieved the best results all-round. 
The second-best model in this task in terms of \textit{positive F1} is Random Forest (\texttt{RF}) that achieved competitive results in all metrics and second-best results in positive class \textit{recall} and overall \textit{micro F1}. Despite the relatively solid performance exhibited by some models (e.g., \texttt{GB} and \texttt{RF}), the results from Table~\ref{tab:results-classification} demonstrate the difficulty of this task. First of all, in general the results obtained by the trained models were encouraging but not exceptionally good, with the best result being \textit{micro F1} $= 0.914$ by \texttt{GB}. Second, the difficulty faced by the trained models is also testified by the small margin by which some of the models ---such as K-Nearest Neighbors (\texttt{KNN}) and Naive Bayes (\texttt{NB})--- outperformed the baselines, with Naive Bayes (\texttt{NB}) even having lower \textit{precision} and \textit{positive F1} with respect to those achieved by \texttt{DT-NFE} in the hard abandonment task. However, for certain models our pipeline proved highly beneficial, significantly surpassing baseline performance as in the case of Support Vector Machine (\texttt{SVM}). Finally, results in table highlight a clear trend where the more complex models consistently outperformed the simpler ones. This shows that the increased complexity of models such as \texttt{GB}, \texttt{RF}, and \texttt{AB} was needed in order to obtain more accurate predictions. 
The scores reported in the bottom half of Table~\ref{tab:results-classification} reveal more nuanced results for the soft abandonment task. Again, the best-performing model is Gradient Boosting (\texttt{GB}), achieving \textit{positive F1} $= 0.590$. Figure~\ref{fig:pr_curves} shows precision-recall curves for this model, for both the hard and soft abandonment tasks. In hard abandonment, precision initially rises sharply and then stabilizes at a relatively high level for a wide range of recall values, before gradually decreasing. Conversely, in soft abandonment, while the initial trend is similar, precision declines more steeply as recall increases. This indicates that as the model attempts to retrieve more positive instances, it misclassifies a greater number of negative instances, leading to a faster deterioration in precision.

A first interesting finding is that the \textbf{overall} results are worse for the soft abandonment task than for hard abandonment, while those for the \textbf{positive class} are better. For example, the best \textit{micro F1} $= 0.814$ in soft abandonment, versus best \textit{micro F1} $= 0.914$ in hard abandonment. 

On the contrary, for the simpler models that underwent our pipeline the best \textit{positive} \textit{F1} $= 0.499$ in soft abandonment, which is higher than \textit{F1} $= 0.434$ in hard abandonment. In other words, by moving from hard to soft abandonment we improved predictions on the positive class (i.e., the abandoning users) but we degraded those on the negative one (i.e., the users who remained active). There are two related reasons for this behavior: (\textit{i}) the definition of abandoning users and (\textit{ii}) class imbalance. For the former, we recall that the soft abandonment task was introduced precisely to address the issue with those users who maintain some activity in the aftermath of the ban, but that eventually leave the platform, which represent challenging instances to classify. Therefore, their labeling in the soft abandonment task as abandoning users (positive class) rather than active ones (negative class), led to a certain degree of improvement of the models on the positive class. The second related factor is about class imbalance given that, before rebalancing, the soft abandonment task is less imbalanced than hard abandonment: 73/27 versus 85/15, respectively. To this end, we recall that abandoning users are the positive and the minority class in both tasks. As such, models trained for detecting soft abandonment train on a relatively larger number of positive instances than those for hard abandonment, hence the better performance on the positive class. 

\subsection{Predicting abandonment at different levels of user activity}
\label{sec:res-activity-levels}
We now delve deeper into the results by analyzing classification performance in relation to the level of user activity on the platform pre-ban. 
Evaluating classification performance based on user activity bears multiple implications, as it allows assessing whether classifiers exhibit varying levels of performance across different segments of users. For example, particularly active users ---who exert the greatest influence on the platform--- are pivotal in shaping community dynamics and behaviors~\citep{zignani2019mastodon,robertson2022uncommon}. Therefore, correctly predicting their activity after an intervention provides moderators and platform administrators with valuable insights for their content management strategies. Conversely, accurately detecting the least active users ---who are more likely to abandon the platform--- enhances the utility of the prediction system by enabling proactive or alternative measures to retain users and mitigate churn~\citep{trujillo2022make}.

\begin{table}[t]
    \small
    \setlength{\tabcolsep}{2pt}
    \centering
    \begin{tabular}{clrclcrrrrr}
        \toprule
        \multicolumn{5}{c}{} && \multicolumn{2}{c}{\textbf{users}} & \multicolumn{3}{c}{\textbf{evaluation metrics}} \\
        \cmidrule(r){7-8}\cmidrule(l){9-10}
        \textbf{task} & \multicolumn{4}{c}{\textbf{level of activity$^\dagger$}} && \textit{pos.} & \textit{neg.} & \textit{positive F1} & \textit{micro F1} \\
        \midrule
            \multirow{5}{*}{\rotatebox[origin=c]{90}{\textbf{hard aban.}}} & \texttt{VH} & 731 $<$ & \textit{n} & && 353 & 2,896 & 0.335 & 0.591 \\
            & \texttt{HI} & 334 $<$ & \textit{n} & $\leq$ 731 && 386 & 2,858 &  \underline{0.726} & \underline{0.929} \\
            & \texttt{ME} & 156 $<$ & \textit{n} & $\leq$ 334 && 432 & 2,801 & \textbf{0.743}  & \textbf{0.941} \\
            & \texttt{LO} & 55 $<$ & \textit{n} & $\leq$ 156 && 464 & 2,783 & 0.263 & 0.871 \\
            & \texttt{VL} & & \textit{n} & $\leq$ 55 && 793 & 2,477 & 0.548 & 0.768 \\
        \midrule
            \multirow{5}{*}{\rotatebox[origin=c]{90}{\textbf{soft aban.}}} & \texttt{VH} & 731 $<$ & \textit{n} & && 764 & 2,485 & 0.493 & \underline{0.813} \\
            & \texttt{HI} & 334 $<$ & \textit{n} & $\leq$ 731 && 742 & 2,502 & 0.448 & 0.481 \\
            & \texttt{ME} & 156 $<$ & \textit{n} & $\leq$ 334 && 759 & 2,474 & 0.536 &  \textbf{0.842} \\
            & \texttt{LO} & 55 $<$ & \textit{n} & $\leq$ 156 && 833 & 2,414 & \underline{0.579}  & 0.794 \\
            & \texttt{VL} & & \textit{n} & $\leq$ 55 && 1,286 & 1,984 & \textbf{0.599} & 0.685 \\
        \bottomrule
        \multicolumn{11}{l}{$\dagger$ \texttt{VH}: Very High; \texttt{HI}: High; \texttt{ME}: Medium; \texttt{LO}: Low; \texttt{VL}: Very Low}
    \end{tabular}
    \caption{Gradient Boosting (\texttt{GB}) classification results in the hard abandonment (top rows) and soft abandonment (bottom rows) tasks, for users with different activity levels. For each activity level, we train a different model and report the corresponding range of comments \textit{n}, the number of users in each class, and the evaluation metrics: \textit{F1 score} of the \textit{positive} class, overall \textit{micro F1}. For each task, the best result in each evaluation metric is shown in \textbf{bold} and the second-best is \underline{underlined}.}
    \label{tab:results-activity}
\end{table}
 
To perform this analysis, we examined the distribution of user comments pre-ban and we assign each user to one of the following five activity levels: very low (\texttt{VL}), low (\texttt{LO}), medium (\texttt{ME}), high (\texttt{HI}), and very high (\texttt{VH}) activity. The five levels are obtained by binning the distribution of user comments at regular intervals of 20 quantiles each, so that \texttt{VL} $\le Q_{20}$, $Q_{20} <$ \texttt{LO} $\le Q_{40}$, and so on up to \texttt{VH} $> Q_{80}$. We then performed a 80/20 train-test split of the dataset stratifying not only for class labels but also for the newly defined activity levels.
Finally, we picked the best performing model based on the results in Table~\ref{tab:results-classification} ---that is, Gradient Boosting (\texttt{GB})--- and we trained and optimized a different model for each activity level. The results of this analysis are reported in Table~\ref{tab:results-activity} for both the hard (top rows) and soft abandonment (bottom rows) task. In table are reported the number of comments \textit{n} in each activity level, the corresponding number of abandoning (positive class) and remaining (negative class) users, and the evaluation metrics: the \textit{F1 score} of the \textit{positive} class, and the overall \textit{micro F1}. The number of users in the positive and negative classes, reported in Table~\ref{tab:results-activity} for each activity level and for each task, shows that the imbalance in the dataset increases when considering increasingly active users. As we already observed in the results of Table~\ref{tab:results-classification}, class imbalance in the data has a detrimental effect on the classification performance on the minority class (i.e., the positive class, corresponding to the abandoning users). In Table~\ref{tab:results-activity}, this is reflected by the fact that the best \textit{positive F1} $= 0.599$ is achieved on the soft abandonment task for users with very low (\texttt{VL}) activity, which is the experimental condition with the lowest class imbalance (61/39). However, this is not true for the hard abandonment task, in which the lowest \textit{positive F1} is obtained for those users whose activity level is low (\texttt{LO}), while the best are obtained for high (\texttt{HI}) and medium (\texttt{ME}) activity levels.
The best overall result is achieved in the hard abandonment task for users with medium (\texttt{ME}) activity, with \textit{micro F1} $= 0.941$. Then, in the soft abandonment task the best overall results are obtained at the medium (\texttt{ME}) and very high (\texttt{VH}) activity levels, with \textit{micro F1} $= 0.842$ and $0.813$, respectively. 

\begin{table}[h]
\centering
\begin{tabular}{cccrrrrrrrrr}
  \toprule
 & \multicolumn{2}{c}{} & \multicolumn{2}{c}{\textbf{Kendall's $\tau$}} & \multicolumn{2}{c}{\textbf{Spearman's $\rho$}} & \multicolumn{2}{c}{\textbf{RBO}} \\  
\cmidrule(r){4-5}\cmidrule(l){6-7}\cmidrule(l){8-9}
 \textbf{task}&   \multicolumn{2}{c}{\textbf{comparisons$^\dagger$}}   & $p = 10$ & $p = \text{all}$ & $p = 10$ & $p = \text{all}$ & $p = 10$ & $p = \text{all}$ \\  
\midrule
  \multirow{6}{*}{\rotatebox[origin=c]{90}{\textbf{hard abandon.}}} & \texttt{VH}     & \texttt{HI}          & $-$0.241 & 0.231 & $-$0.311 & 0.580 & 0.280  & 0.420 \\   
& \texttt{HI}          & \texttt{ME}        & $-$0.009 & 0.022 & $-$0.329 & 0.323 & 0.258  & 0.389 \\   
& \texttt{ME}        & \texttt{LO}           & $-$0.222 & 0.018 & $-$0.263 & 0.296 & 0.245  & 0.375 \\   
& \texttt{LO}           & \texttt{VL}     & $-$0.205 & 0.128 & $-$0.415 & 0.239 & 0.235  & 0.340 \\  
\cmidrule(l){2-9} 
& \texttt{VH}     & \texttt{VL}      & $-$0.205 & 0.231 & $-$0.291 & 0.135 & 0.250  & 0.295 \\ 
& \texttt{HI}         & \texttt{LO}         & $-$0.402 & 0.005 & $-$0.329 & $-$0.272 & 0.258  & 0.310 \\   
\midrule
  \multirow{6}{*}{\rotatebox[origin=c]{90}{\textbf{soft abandon.}}}& \texttt{VH}     & \texttt{HI}          & $-$0.352 & 0.128 & $-$0.178 & 0.191 & 0.239  & 0.389 \\   
& \texttt{HI}         & \texttt{ME}        & $-$0.188 & 0.005 & $-$0.329 & $-$0.014 & 0.267  & 0.357 \\   
& \texttt{ME}        & \texttt{LO}           & $-$0.365 & 0.018 & $-$0.263 & $-$0.013 & 0.267  & 0.342 \\   
& \texttt{LO}           & \texttt{VL}      & $-$0.120 & 0.128 & $-$0.415 & $-$0.038 & 0.235  & 0.330 \\   
\cmidrule(l){2-9}
& \texttt{VH}     & \texttt{VL}      & $-$0.239 & 0.231 & $-$0.291 & 0.191 & 0.250  & 0.307 \\   
& \texttt{HI}          & \texttt{LO}           & $-$0.402 & 0.005 & $-$0.329 & $-$0.272 & 0.258  & 0.310 \\   
\bottomrule
\multicolumn{9}{l}{$\dagger$ \texttt{VH}: Very High; \texttt{HI}: High; \texttt{ME}: Medium; \texttt{LO}: Low; \texttt{VL}: Very Low}
\end{tabular}
\caption{Agreement between the feature rankings of different models, each corresponding to users with a specific activity level. The \textbf{comparisons} column indicates which models are compared. Agreement between feature rankings is measured in terms of Kendall's $\tau$ and Spearman's $\rho$ rank correlations, and via the Rank-Biased Overlap (RBO) ranking similarity. Feature rankings are compared by considering all features ($p = \text{all}$) or only the top-10 most important ones ($p = 10$).}
\label{tab:correlations}
\end{table}
 
After training a different model for each activity level, we assessed whether these models assigned similar importance to the features they used.
We operationalized this analysis by computing several indicators of correlation and similarity between the feature rankings of the different models, such as Kendall's $\tau$, Spearman's $\rho$ and Rank-Biased Overlap (RBO)~\citep{webber2010similarity}. To rank the features of a model, we first identified and discarded strongly correlated features. In detail, we computed Pearson correlations \textit{r} between each feature pair of a model, we identified those feature pairs for which $\textit{r} > |0.7|$, and we discarded one of such features at random. Then, for each model, we computed the feature importance of the remaining (uncorrelated) features with Permutation Importance (PI). PI measures the contribution of each feature to the model's predictions by quantifying the performance degradation observed when its values are randomly shuffled while keeping all other features unchanged. Results are presented in Table \ref{tab:correlations} for the top-10 features of each model ($p = 10$) and for all features ($p = \text{all}$). The column \textbf{comparisons} indicates which models ---each corresponding to a different user activity level--- are compared. As reported, we compared the feature ranking of models corresponding to similar activity levels (e.g., \texttt{VH} \textit{vs.} \texttt{HI}), as well as those corresponding to opposite ones (e.g., \texttt{VH} \textit{vs.} \texttt{VL} and \texttt{HI} \textit{vs.} \texttt{LO}). The underlying intuition is that similar rankings should emerge among similar activity levels, whereas opposite levels are expected to exhibit less agreement.

For the hard abandonment task, results in Table \ref{tab:correlations} show that considering all features led to higher agreement, suggesting that the correlation tended to increase towards the tail of the rankings rather than the head. For instance, the correlation between \texttt{VH} and \texttt{HI} showed strong agreement, with Spearman’s $\rho = 0.580$ and RBO $= 0.420$ when $p = \text{all}$, but decreased when $p = 10$. This suggests that highly active users share similar characteristics and can be grouped together. For the adjacent levels, this agreement decreases, particularly when transitioning from \texttt{HI} to \texttt{ME} and further down the activity scale. For instance, \texttt{HI} and \texttt{ME} showed a noticeable drop in agreement, with Kendall’s $\tau = -0.009$, indicating that the ranking of features differs significantly. This suggests that while highly active users (\texttt{VH} and \texttt{HI}) may exhibit similar behaviors, comparing similar but lower ranges of activity levels introduces more variability. The agreement between adjacent levels, such as \texttt{HI} and \texttt{ME} or \texttt{ME} and \texttt{LO}, was notably weaker, especially in Kendall’s $\tau$, which remains close to zero or even negative. Surprisingly, the agreement between \texttt{VH} and \texttt{VL} is stronger than some adjacent-level comparisons, with Spearman’s $\rho = 0.135$ and RBO $= 0.295$. While this correlation is still moderate, it suggests that these two groups, despite their opposing activity levels, might share some behavioral patterns. A possible explanation is that both represent users with extreme behaviors, leading to similar feature relevance.

Results in the soft abandonment task present the same patterns of the hard abandonment task. Higher agreement was generally found when all features are considered, with stronger correlations towards the tail of the rankings. For example, \texttt{VH} and \texttt{HI} exhibited relatively high agreement, particularly with respect to RBO, while Kendall’s $\tau$ remained lower, again highlighting sensitivity to rank shifts. Conversely, adjacent levels such as \texttt{HI} and \texttt{ME}, or \texttt{ME} and \texttt{LO}, show weak or even negative rank correlations, reinforcing the idea that mid-range activity users display more behavioral variability. Again, \texttt{VH} and \texttt{VL} show a correlation pattern similar to that of the hard abandonment task, with Spearman’s $\rho = 0.191$ and RBO $= 0.307$. This further supports the idea that extreme user groups, despite their contrasting activity levels, might share unexpected similarities. Overall, while some user groups exhibit similar feature importance rankings, others differ significantly. This highlights the need for tailored approaches rather than \textit{one-size-fits-all} solutions. Understanding which users can be moderated similarly and which require separate treatment is pivotal for developing more personalized and effective moderation strategies~\citep{cresci2022personalized}.

\subsection{Leave-one-out cross-validation}
Until now, we trained our models on a dataset obtained by merging users from all 15 banned subreddits. However, users who participate in different subreddits could have different characteristics and exhibit different behaviors and reactions to a moderation intervention~\citep{trujillo2023one}. For this reason, we carried out an additional analysis aimed at assessing possible subreddit-specific factors influencing user abandonment. Importantly, this analysis also allows evaluating the extent to which a model trained on a certain set of subreddits is capable of generalizing to other ---unseen--- subreddits. To assess the generalizability of our best model we performed a leave-one-out cross-validation (LOOCV) by leveraging the multiplicity of banned subreddits in our dataset. This procedure represents the state-of-the-art in evaluating the generalizability of a classifier against multiple groups or classes of data instances~\citep{wong2015performance}, and has been recently used for related tasks such as the detection of different groups of bots~\citep{echeverri2018lobo,mannocci2022mulbot}. Based on the 15 banned subreddits in our dataset, we implemented the LOOCV by iteratively selecting 14 subreddits to train a Gradient Boosting (\texttt{GB}) hard and soft abandonment detection classifier, and by testing the trained model on users from the remaining (held-out) subreddit. For this analysis, we considered a user to participate in a subreddit if it posted more than 10 comments in that subreddit, so as to reduce noise caused by sporadic participation.

\begin{figure}
  \centering
\subfloat{\includegraphics[width=0.6\columnwidth]{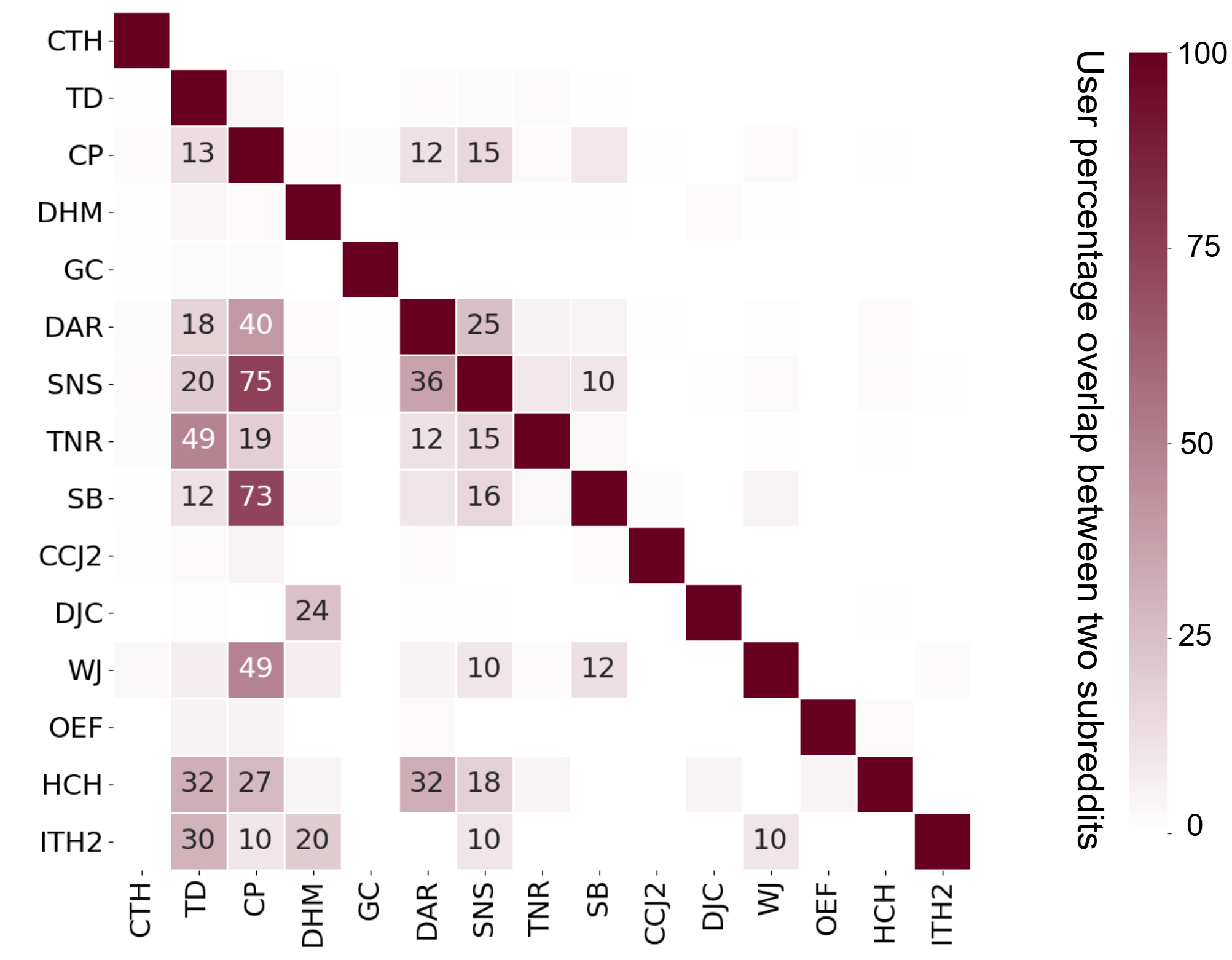} }
  \caption{User overlap matrix $O$ between the banned subreddits. Each cell $o_{ij}$ in the matrix reports the percentage of users from the $i$-th subreddit that also participated in the $j$-th subreddit. White-colored cells represent 0\% overlap while dark-colored ones represent 100\% overlap. The exact amount of overlap is shown for cells with $o_{ij} > 10\%$. Subreddits are ordered by decreasing number of participating users.}\label{fig:overlap}
\end{figure}

The validity of the LOOCV hinges on the assumption of independence between the training and testing datasets. Specifically, to ensure unbiased evaluation, it is required that the held-out subreddit contains users who are largely distinct from those in the training subreddits. While the independence assumption is easily verified in some domains of application, such as that of bot detection where each bot belongs to only one botnet~\citep{echeverri2018lobo}, it does not hold in our context where any user can participate in multiple subreddits. 
When a considerable overlap exists between the sets of users who participate in two subreddits, utilizing either of these subreddits as the held-out test dataset may leak information to the classifier from the training datasets, thereby compromising the validity of the evaluation~\citep{cresci2024demystifying}. Before running the LOOCV we thus assessed the extent of overlap between the sets of users who participated in the 15 banned subreddits. The results of this analysis are presented in Figure~\ref{fig:overlap} as a heatmap of the user overlap matrix between the subreddits. In figure, each cell $o_{ij}$ in the user overlap matrix $O$ reports the percentage of users from the $i$-th subreddit that also participated in the $j$-th subreddit. By definition, cells in the matrix diagonal correspond to 100\% overlap and are dark-colored, while those that correspond to no overlap are white colored. Apart from \subr{ShitNeoconsSay} (SNS) and \subr{soyboys} (SB) that share up to 75\% of their users with \subr{ConsumeProducts} (CP), the matrix is overall very sparse, as demonstrated by the limited reported overlap. The results from this analysis show that the sets of users who participated in the banned subreddits are largely disjoint, in line with the literature on echo chambers~\citep{cinelli2021echo}, which supports the application of the LOOCV.

\begin{table*}[t]
    \small
    \centering
    \setlength{\tabcolsep}{2.5pt}
    \begin{tabular}{cllrrcrrrcrrr}
        \toprule
        \multicolumn{5}{c}{} && \multicolumn{3}{c}{\textbf{positive class}} && \multicolumn{2}{c}{\textbf{overall}}\\ 
        \cmidrule{7-9}\cmidrule{11-12}
        \textbf{task} & \multicolumn{2}{l}{\textbf{test subreddit}} & \textbf{users} & \textbf{max overlap} && \textit{precision} & \textit{recall} & \textit{F1} &&  \textit{AUC} &  \textit{micro F1} \\
        \midrule
            \multirow{17}{*}{\rotatebox[origin=c]{90}{\textbf{hard abandonment}}} & CTH & \subr{ChapoTrapHouse} & 8,471 & 0.2\% && .111 & .859 & .197 && .685 & .367  \\
            & TD     & \subr{The\_Donald} & 3,539 & 4.1\% && .801 & .514 & .626  && .892 & .855 \\
            & CP     & \subr{ConsumeProduct} & 1,098 & 15.0\% && .806 & .083 & .151 && .720 & .741 \\
            & DHM    & \subr{DarkHumorAndMemes} & 929 & 3.8\% && .737 & .625 & .676 && .931 & .926\\
            & GC     & \subr{GenderCritical} & 918 & 1.4\% && .548 & .687 & .610  && .764 & .708\\
            & DAR    & \subr{DebateAltRight} & 325 & 40.0\% && .769 & .100 & .177 && .615 & .697 \\
            & SNS    & \subr{ShitNeoconsSay} & 226 & 75.0\% && .556 & .851 & .673 &&  .808 & .676\\
            & TNR    & \subr{TheNewRight} & 144 & 49.0\% && .000 & .000 & .000  && .544 & .789\\
            & SB     & \subr{soyboys} & 142 & 73.0\% && .800 & .190 & .308 &&  .738 & .746\\
            & CCJ2   & \subr{CCJ2} & 110 & 4.5\% && .600 & .882 & .714 && .961 & .889\\
            & DJC    & \subr{DarkJokeCentral} & 94 & 24.0\%  && 1.000 & .100 & .182 && .587 & .899\\
            & WJ     & \subr{Wojak} & 57 & 49.0\% && .846 & .846 & .846 && .963 & .927 \\
            & OEF    & \subr{OandAExclusiveForum} & 37 & 5.4\% && .286 & 1.000 & .444 && .914 & .865\\
            & HCH    & \subr{HateCrimeHoaxes}& 22 & 32.0\% && .714 & .500 & .588 && .767 & .682\\
            & ITH2   & \subr{imgoingtohellforthis2} & 10 & 30.0\% && .000 & .000 & .000 && .875 & .889\\
            \cmidrule{2-12}
            & \multicolumn{2}{l}{averaged LOOCV results} &&&& \makecell[r]{.571 \\ $\pm$ .311} & \makecell[r]{ .482\\ $\pm$ .356} & \makecell[r]{.412\\ $\pm$ .271} &&  \makecell[r]{ .784\\ $\pm$ .132} & \makecell[r]{ .777\\ $\pm$.141}\\
            & \multicolumn{2}{l}{training/testing on all subreddits (no LOOCV)} &&&& .658 &  .736 & .695 && .930 & .914\\
        \midrule
            \multirow{17}{*}{\rotatebox[origin=c]{90}{\textbf{soft abandonment}}} & CTH & \subr{ChapoTrapHouse} & 8,471 & 0.2\% && .184 & .892 & .305 && .551 & .250\\
            & TD     & \subr{The\_Donald} & 3,539 & 4.1\% &&  .454 & .712 & .554 && .608 & .550\\
            & CP     & \subr{ConsumeProduct} & 1,098 & 15.0\% && .470 & .582 & .520 && .517 & .516\\
            & DHM    & \subr{DarkHumorAndMemes} & 929 & 3.8\% && .367 & .648 & .469 && .680 & .621\\
            & GC     & \subr{GenderCritical} & 918 & 1.4\% && .627 & .747 & .682 && .744 & .672\\
            & DAR    & \subr{DebateAltRight} & 325 & 40.0\% && .599 & .617 & .608 && .553 & .567\\
            & SNS    & \subr{ShitNeoconsSay} & 226 & 75.0\% && .598 & .869 & .708 && .570 & .581\\
            & TNR    & \subr{TheNewRight} & 144 & 49.0\% && .443 & .879 & .590 && .525 & .500\\
            & SB     & \subr{soyboys} & 142 & 73.0\% &&   .496 & .896 & .638 && .536 & .521\\
            & CCJ2   & \subr{CCJ2} & 110 & 4.5\% && .346 & .848 & .491 && .528 & .463 \\
            & DJC    & \subr{DarkJokeCentral} & 94 & 24.0\% && .204 & .733 & .319 && .607 & .472\\
            & WJ     & \subr{Wojak} & 57 & 49.0\% && .595 & .880 & .710 && .624 & .673\\
            & OEF    & \subr{OandAExclusiveForum} & 37 & 5.4\% && .217 & .714 & .333 && .695 & .459\\
            & HCH    & \subr{HateCrimeHoaxes} &22 & 32.0\% && .556 & .833 & .667 &&  .425 & .545\\
            & ITH2   & \subr{imgoingtohellforthis2} & 10 & 30.0\%  && .333 & .667 & .444 && .500 & .444\\
            \cmidrule{2-12}
            & \multicolumn{2}{l}{averaged LOOCV results} &&&& \makecell[r]{ .432 \\ $\pm$ .146} & 	\makecell[r]{.767 \\ $\pm$ .105} & \makecell[r]{ .535 \\ $\pm$.135 } &&  \makecell[r]{ .535 \\ $\pm$.135 } & \makecell[r]{ .522\\ $\pm$ .100}\\
            & \multicolumn{2}{l}{training/testing on all subreddits (no LOOCV)} &&&& .513 & .694 & .590 && .816  &.808 \\
        \bottomrule
    \end{tabular}
\caption{Results of the leave-one-out cross-validation (LOOCV) analysis. The top half of the table reports results for the hard abandonment task, while the bottom half is related to the soft abandonment. For both tasks, each row shows the classification results obtained with the reported subreddit used as test set and the remaining ones as training set. For each task, the bottom rows report the LOOCV results aggregated over all subreddits (mean $\pm$ standard deviation) and the reference results obtained by the same model without the LOOCV.}
\label{tab:results-loocv}
\end{table*}
 
Table~\ref{tab:results-loocv} shows the results of the LOOCV for the hard (top rows) and soft abandonment (bottom rows) tasks. Each row in the table reports testing results on a specific subreddit, when training on all others. In addition to the evaluation metrics, for each subreddit we also report the number of participating users and its maximum overlap with the other subreddits. Finally, for each task, the bottom rows report evaluation results aggregated across all subreddits, in terms of mean and standard deviation of each metric, as well the scores obtained by the \texttt{GB} model trained and tested on data from all subreddits ---that is, without the LOOCV--- as reported in Table~\ref{tab:results-classification}. The latter scores serve as reference values with which to compare the LOOCV results to quantify the performance decrease to expect when classifying users from an unseen subreddit. The averaged results reported in Table~\ref{tab:results-loocv} for both tasks indicate a high variability in the evaluation metrics of the positive class, as demonstrated by the large standard deviation with respect to the means. For instance, in the hard abandonment task, the \textit{F1 score} ranges from a minimum of $0.000$ for \subr{TheNewRight} (TNR) and \subr{imgoingtohellforthis2} (ITH2) to a maximum of $0.846$ for \subr{Wojak} (WJ). Similarly, the \textit{recall} ranges from $0.000$ for \subr{TheNewRight} (TNR) and \subr{imgoingtohellforthis2} (ITH2) to $1.000$ for \subr{OandAExclusiveForum} (OAE). The same can be observed for the soft abandonment task as well, where the \textit{precision} ranges from $0.184$ for \subr{ChapoTrapHouse} (CTH) to $0.827$ for \subr{GenderCritical} (GC). 
These results are indicative of substantial differences among the banned subreddits. In fact, in addition to the overall limited user overlap observed in Figure~\ref{fig:overlap}, the results from Table~\ref{tab:results-loocv} highlight notable differences in the behavior exhibited by the participants of such subreddits. 
Interestingly enough, the maximum overlap that each subreddit has with others, has little influence on the performance of the trained model. For each task, we computed the Pearson correlation coefficient \textit{r} to estimate the strength of the linear relationship between the max overlap and two evaluation metrics ---namely, the \textit{F1 score} of the positive class and the \textit{micro F1}. The largest value measured for the hard abandonment task is \textit{r} $= 0.237$ (\textit{p} $=0.394$), while the largest measured for the soft abandonment task is \textit{r} $= 0.222$ (\textit{p} $=0.493$), both of which indicate weak and non-significant correlations.
In turn, this strengthens the results reported in Table~\ref{tab:results-loocv}, including those about subreddit diversity, as the variable results in the table cannot be simply explained by the fact that some users participated in multiple subreddits, but rather to their inherent differences. The comparison between the results obtained without LOOCV and the LOOCV aggregate results sheds light on the capacity of our model to generalize to unseen data. To this end, results in Table~\ref{tab:results-loocv} show a moderate loss in performance in the LOOCV experiment, which is expected given the aforementioned differences among the subreddits. However, the percentage loss is generally contained. For example, in the hard abandonment task the average \textit{F1 score} on the positive class obtained with the LOOCV is 60.0\% of the original one.  The loss for the overall metrics is even lower, with the average \textit{micro F1} reaching 85.0\% of the original value. Similar percentage losses are obtained for the soft abandonment task in the overall metrics, with the LOOCV average \textit{positive F1} maintaining 90.7\% of the original value. Overall, these figures suggest that the model maintains encouraging performance even when applied to unseen data.

\begin{figure*}[t]
    \begin{subfigure}[t]{0.49\columnwidth}\centering
        \includegraphics[width=\columnwidth]{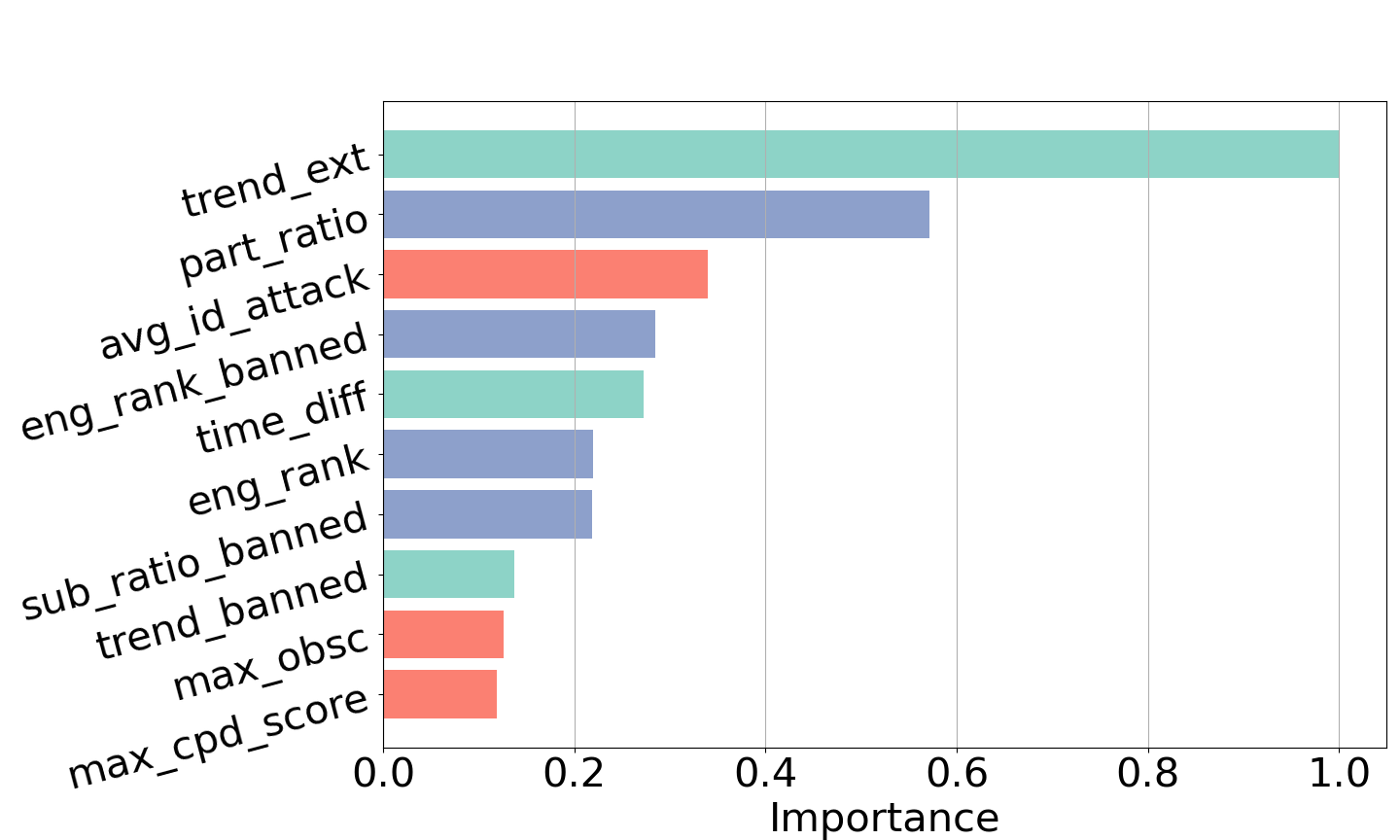}
        \caption{Task: hard abandonment. Feature importance: SHAP.}
        \label{fig:imp_hard}\end{subfigure}
    \begin{subfigure}[t]{0.49\columnwidth}\centering
        \includegraphics[width=\columnwidth]{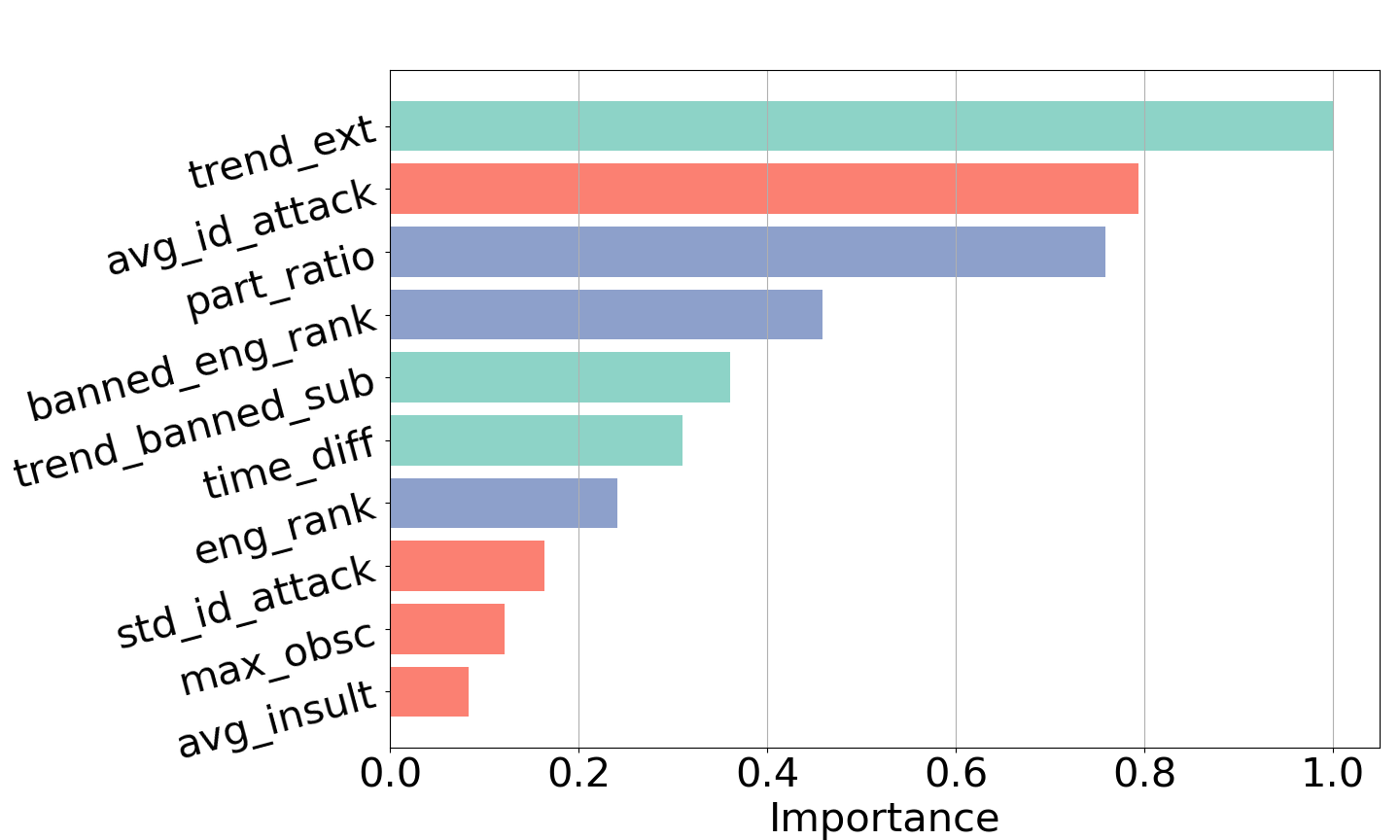}
        \caption{Task: soft abandonment. Feature importance: SHAP.}
        \label{fig:imp_soft}\end{subfigure}
    \par\bigskip
    \begin{subfigure}[t]{0.49\columnwidth}\centering
        \includegraphics[width=\columnwidth]{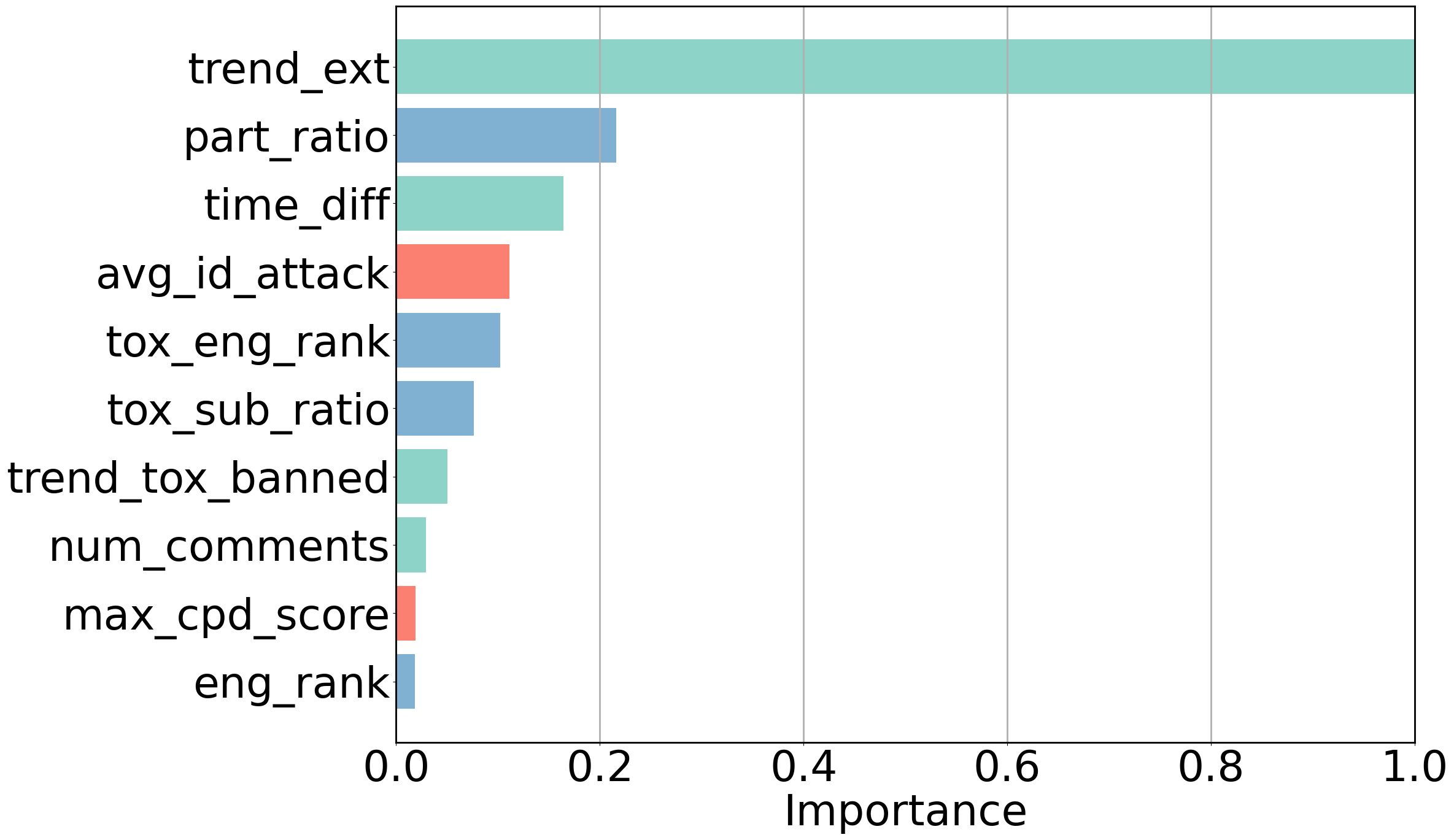}
        \caption{Task: hard abandonment. Feature importance: PI.}
        \label{fig:pi_imp_hard}\end{subfigure}
    \begin{subfigure}[t]{0.49\columnwidth}\centering
        \includegraphics[width=\columnwidth]{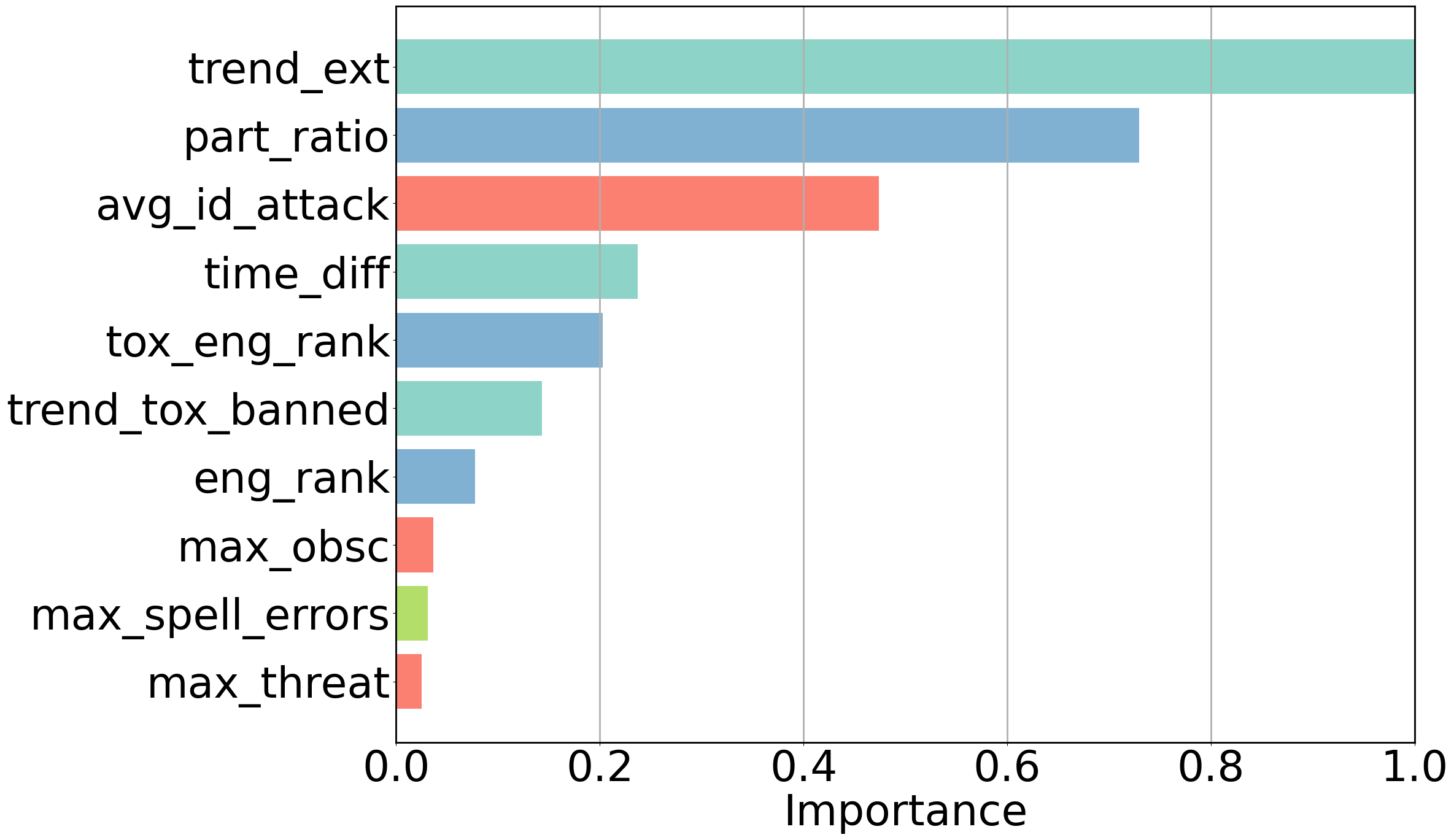}
        \caption{Task: soft abandonment. Feature importance: PI.}
        \label{fig:pi_imp_soft}\end{subfigure}
    \caption{Top-10 most important features for the hard and soft abandonment tasks according to \textbf{SHAP} (top row) and \textbf{PI} (bottom row). The ranking of the most important features is similar between the two tasks, with features such as \textit{trend\_ext}, \textit{part\_ratio}, and \textit{avg\_id\_attack} dominating the ranking. All classes of features are extensively represented in the top-10, with the exception of \textit{writing style} features that do not seem to provide individually strong contributions to model predictions.}
    \label{fig:imp_combined}
\end{figure*}

\subsection{Feature importance}
\label{sec:feat-imp}
Our next analysis involves estimating the importance of the individual features as well as of the classes of features that we implemented, for both the hard and soft abandonment tasks. This analysis complements our previous results by providing additional information for interpreting model predictions and by highlighting which features are most predictive of user abandonment following a moderation intervention. Not only does such analysis enhance the interpretability of our models, but it also provides moderators with valuable insights into the key factors of user abandonment. By understanding which features contribute the most to the predicted outcomes, moderators would be able to identify patterns and refine their moderation strategies accordingly. In general, transparency can help moderators make better informed decisions, balancing rule enforcement with user retention, and ensuring that moderation remains effective and fair.

We rely on SHapley Additive exPlanations (SHAP) and Permutation Importance (PI) as the building blocks of our feature importance analysis. PI was introduced in Section~\ref{sec:res-activity-levels}. Instead, SHAP is a method based on cooperative game theory that is widely used for interpreting the output of machine learning models and for estimating the contribution of individual features to a model's predictions~\citep{lundberg2017unified,mazza2022investigating}.
Leveraging two different post-hoc explainers allows us to gain a deeper understanding of the importance of the implemented features for this task.

To obtain the overall importance of each feature for our task, we use an information fusion-based sensitivity analysis technique~\citep{tardelli2022detecting}. Concretely, SHAP estimates feature importance at the individual data point level, requiring values to be aggregated across the entire dataset. 
Therefore, for each feature $f_i$ and for each trained model $M_j \in \{\texttt{NB},\texttt{KNN},\texttt{DT},\texttt{RF},\texttt{AB},\texttt{GB},\texttt{SVM}\}$, we use SHAP to compute the contribution $C(i,j,d)$ of $f_i$ to the prediction given by $M_j$ for a specific data point $d$. We then aggregate the contributions over all data points $N$ in the test set to compute a local score: \begin{equation}
\text{score}_L(i,j) = \frac{1}{N}\sum_{d=1}^N\big|C(i,j,d)\big|.
\end{equation}
The $\text{score}_L(i,j)$ expresses the \textit{local} contribution ---or importance--- of the feature $f_i$ for the model $M_j$. Next, we obtain a \textit{global} score for the feature $f_i$ as the weighted mean of its local importance across all trained models:
\begin{equation}
\label{eq:global-score}
\text{score}_G(i) = \frac{\sum_j w_j \cdot \text{score}_L(i,j)}{\sum_j w_j}.
\end{equation}
We use the \textit{F1 score} on the positive class as the weighting factor $w_j$ of each model $M_j$, so that models that are better at detecting abandoning users are weighted more in Eq.~\eqref{eq:global-score}. The $\text{score}_G(i)$ expresses the weighted contribution of the feature $f_i$ throughout all models that we trained and tested ---hence the \textit{global} label--- thus representing a robust indicator of the actual importance of that feature for the task. Contrarily to SHAP, PI estimates feature importance for whole models, rather than for individual data points. Thus, when computing feature importance with PI we directly apply Eq.~\eqref{eq:global-score} to obtain the global feature importance scores. Before that, we remove correlated features (i.e., those with Pearson $r > |0.7|$), as already done in Section~\ref{sec:res-activity-levels}, to avoid possible misleading results. As a final presentation step, we rank all features and assess their relative importance by normalizing their global scores so that the best performing feature has a normalized score of 1 and the others are rescaled proportionally.

Figure~\ref{fig:imp_combined} shows the top-10 most important features according to their normalized global scores, for the hard and soft abandonment tasks, with feature importance computed via SHAP and PI. The set of the most important features for both tasks and explainers is very similar, although the ranking is different and the features exhibit different relative importance. In both tasks, among the most informative features are the trend of comments in non-banned subreddits (\textit{trend\_ext}), the ratio between the number of non-banned and banned subreddits to which a user participated (\textit{part\_ratio}), the average score of ``identity attacks'' in the user's comments (\textit{avg\_id\_attack}), and the average time passed between the post of two subsequent comments (\textit{time\_diff}). For both tasks and explainers, the single most informative feature is by far \textit{trend\_ext}. To this end we recall that the \texttt{DT Trend} baseline defined in Section~\ref{sec:experiments-comparisons} is based on this exact feature, which explains the good performance of that baseline in Table~\ref{tab:results-classification}. 
We also note that in both tasks the first few features provide a much larger relative contribution than the last ones in the top-10. This means that, apart from the first few features, all others, including those highly ranked according to our feature importance analysis, provide relatively few information to the models. This result resonates with that presented in Table~\ref{tab:results-classification} about the small number of features used by non-ensemble models. Finally, we note that three out of four classes of features are extensively represented among the top-10 features shown in Figure~\ref{fig:imp_combined}. The only class of features that did not make it to the top-10 ---except for the 9th ranked features according to PI for the soft abandonment task--- is the one conveying information the \textit{writing style} of the users, which does not seem to provide much relevant information.

\begin{figure*}[t]
    \begin{subfigure}[t]{0.49\columnwidth}\centering
        \includegraphics[width=\columnwidth]{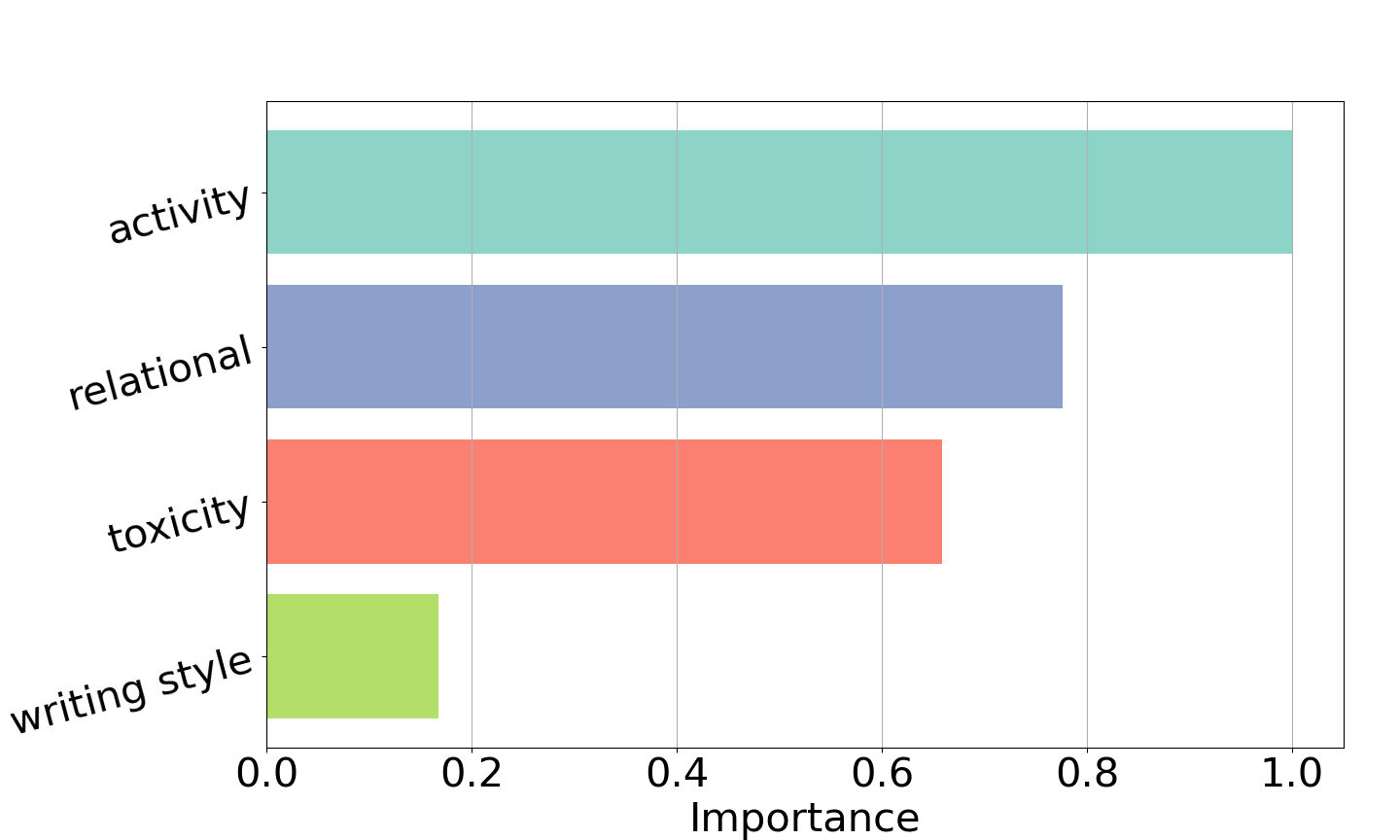}
        \caption{Task: hard abandonment. Feature importance: SHAP.}
        \label{fig:imp_hard_class}\end{subfigure}
    \begin{subfigure}[t]{0.49\columnwidth}\centering
        \includegraphics[width=\columnwidth]{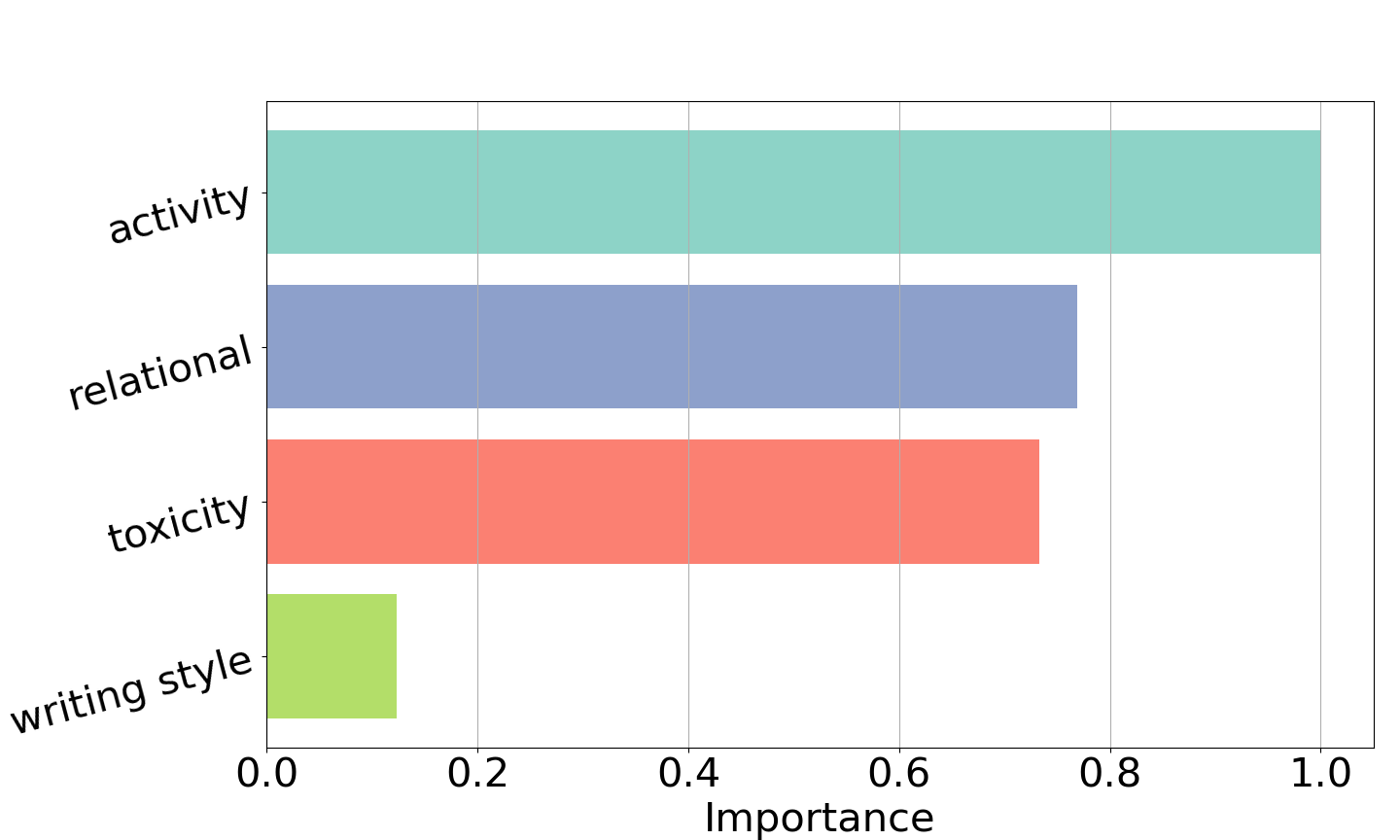}
        \caption{Task: soft abandonment. Feature importance: SHAP.}
        \label{fig:imp_soft_class}\end{subfigure}
    \par\bigskip
    \begin{subfigure}[t]{0.49\columnwidth}\centering
        \includegraphics[width=\columnwidth]{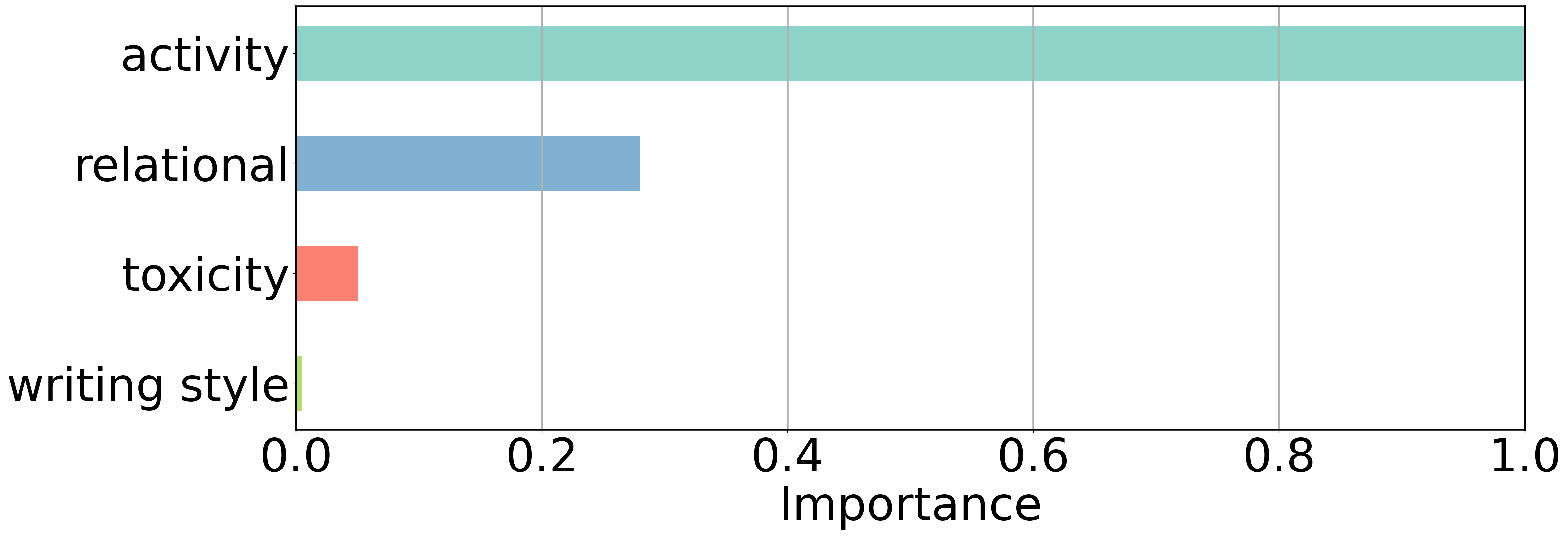}
        \caption{Task: hard abandonment. Feature importance: PI.}
        \label{fig:pi_imp_hard_class}\end{subfigure}
    \begin{subfigure}[t]{0.49\columnwidth}\centering
        \includegraphics[width=\columnwidth]{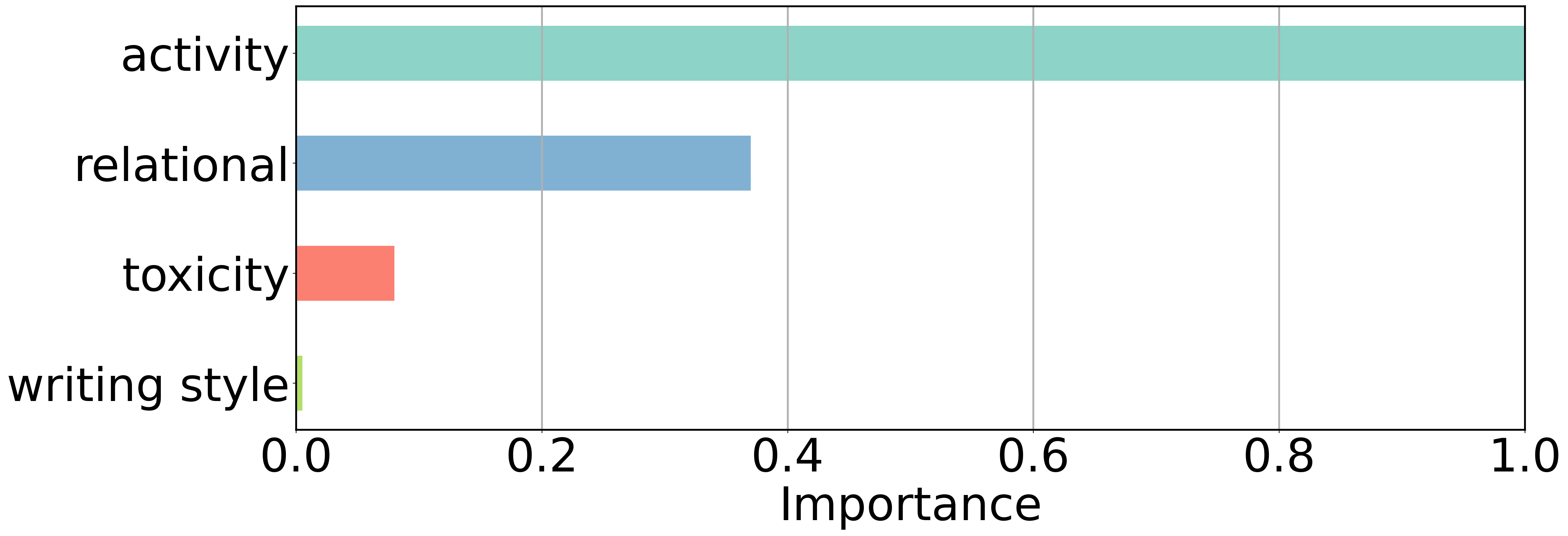}
        \caption{Task: soft abandonment. Feature importance: PI.}
        \label{fig:pi_imp_soft_class}\end{subfigure}
    \caption{Feature importance for the hard and soft abandonment tasks according to \textbf{SHAP} (top row) and \textbf{PI} (bottom row). The importance is calculated for each class of features by summing the contributions of individual features in each class, normalized by the number of features in that class. The contributions of the different classes of features are largely the same across the two tasks, with \textit{activity} features providing the largest contribution, followed by \textit{relational} and \textit{toxicity} features. \textit{Writing style} features provide overall marginal contributions.}
    \label{fig:imp_combined_class}
\end{figure*}

To better assess the contribution of the different classes of features, rather than that of the individual features, we aggregate the contributions of all features based on their class. In detail, we sum the global importance scores of all features separately for each class, and we divide it by the number of features in the class, to obtain a class feature importance score. Lastly, we normalize the class scores so that the best class has a normalized score of 1 and the others are rescaled proportionally. Figure~\ref{fig:imp_combined_class} shows the relative importance of the different classes of features for the two tasks according to SHAP and PI. The ranking is the same for both tasks and explainers, with \textit{activity} features providing the largest contribution.
\textit{Relational} features are the second-most informative class, providing about 80\% of the contribution of activity features. \textit{Toxicity} features obtain a similar score. Instead, \textit{writing style} features are ranked last in both tasks and provide small contributions with respect to the other classes. This result confirms that shown in Figure~\ref{fig:imp_combined}, where no \textit{writing style} feature was ranked among the top contributing features in either task. Overall, the results obtained for \textit{activity}, \textit{relational}, and \textit{toxicity} features highlight the importance of user engagement, social dynamics, and toxic speech in predicting user abandonment after the investigated moderation intervention. Conversely, information about the \textit{writing style} of the users does not appear to be as relevant.

\subsection{Class imbalance}
\begin{table}[t]
    \small
    \centering
    \setlength{\tabcolsep}{2.5pt}
    \begin{tabular}{clcccccc}
        \toprule
        \multicolumn{3}{c}{} & \multicolumn{3}{c}{\textbf{positive class}} & \multicolumn{2}{c}{\textbf{overall}} \\ 
        \cmidrule(r){4-6}\cmidrule(l){7-8}
        \textbf{task}  & \textbf{rebalancing strategy$^\dagger$} & \textit{p} & \textit{precision} & \textit{recall} & \textit{F1} &  \textit{AUC}  & \textit{micro F1} \\
        \midrule
            \multirow{7}{*}{\rotatebox[origin=c]{90}{\textbf{hard abandonment}}} 
            & --  & 50  & 0.002  & \underline{0.500}  & 0.004  & \underline{0.850}  & 0.738 \\ 
            & Resampling &50 & 0.520 & 0.373 & \textbf{0.434}  & 0.781 & \textbf{0.797} \\
            & Cost-Sensitive Learning  & 60  & \underline{0.679}  & 0.303  & 0.419  & 0.719  & \underline{0.776} \\ 
            & Resampling + Cost-Sensitive Learning  & 40  & \textbf{0.689}  & 0.303  & \underline{0.421}  & 0.717  & 0.769 \\ 
            & Ensemble Learning & 60  & 0.107  & \textbf{0.542}  & 0.179  & \textbf{0.853}  & 0.770 \\
            & Ensemble Learning + Resampling  & 70  & 0.440  & 0.324  & 0.373  & 0.779  & 0.742 \\
            & Ensemble Learning + Cost-sensitive Learning  & 20  & 0.107  & 0.464  & 0.174  & 0.848  & 0.747 \\
        \midrule
            \multirow{7}{*}{\rotatebox[origin=c]{90}{\textbf{soft abandonment}}} 
            & --  & 40  & 0.024  & \underline{0.583}  & 0.046  & 0.732  & \textbf{0.779} \\
            & Resampling & 50 & 0.484 & 0.513 & 0.499 & 0.728 & \underline{0.737} \\
            & Cost-sensitive Learning  & 60  & \underline{0.613}  & 0.440  & \underline{0.513}  & 0.685  & 0.721 \\
            & Resampling + Cost-Sensitive Learning  & 60  & \textbf{0.639}  & 0.449  & \textbf{0.528}  & 0.691  & 0.726 \\
            & Ensemble Learning & 80  & 0.161  & \textbf{0.592}  & 0.253  & \textbf{0.744}  & 0.719 \\
            & Ensemble Learning + Resampling  & 80  & 0.485  & 0.486  & 0.485  & 0.723  & 0.725 \\
            & Ensemble Learning + Cost-sensitive Learning  & 10  & 0.047  & 0.569  & 0.086  & \underline{0.733}  & 0.583 \\
        \bottomrule
        \multicolumn{5}{l}{$\dagger$ Resampling = SMOTE + Random Undersampling}
    \end{tabular}
    \caption{Support Vector Machine (SVM) classification results in the hard abandonment (top rows) and soft abandonment (bottom rows) tasks, obtained by using different rebalancing strategies. Column \textit{p} reports the number of features used by each model. For each task, the best result in each evaluation metric is shown in \textbf{bold} and the second-best is \underline{underlined}.}
    \label{tab:Resamplingults-imbalance}
\end{table}
 Motivated by the observed degrading effect of class imbalance, we trained multiple Support Vector Machines (\texttt{SVM}) to experiment with a few different rebalancing strategies, used both in isolation and in combination. 
Among the strategies that we considered are SMOTE, Random Undersampling, Cost-Sensitive Learning, and Ensemble Learning. In addition, we also trained a classifier on the natural class distribution, without carrying out any rebalancing. We discussed the use of SMOTE and Random Undersampling in Section~\ref{sec:exp-scaling-rebalancing}. Cost-Sensitive Learning balances the class weights by adjusting the loss function so as to increase the penalty for misclassifying the minority class, thus making the model pay more attention to correctly identifying those minority instances. Finally, the Ensemble Learning strategy implements a majority-voting ensemble by training ten classifiers on the dataset with a non-deterministic process, and then combining their outputs via majority voting.

Table \ref{tab:Resamplingults-imbalance} reports the results of these analyses. For both tasks, the results clearly show that the \texttt{SVM} without rebalancing  is heavily biased toward the majority class, leading to poor performance on the positive class. Moreover, while using an ensemble of unbalanced \texttt{SVMs} helps improve the performances, the improvement remains limited and it does not significantly mitigate the bias unless resampling is applied beforehand (Ensemble Learning + Resampling). Among all strategies, Resampling and Cost-Sensitive Learning proved to be the most effective in addressing class imbalance, even when combined. In particular, the best result in terms of \textit{positive F1} is obtained with Resampling for the hard abandonment task ($F1 = 0.434$), and with Resampling + Cost-Sensitive Learning for the soft abandonment task ($F1 = 0.528$). These findings highlight the substantial impact of class imbalance on model performance, reinforcing the importance of carefully selecting strategies to mitigate its effects. The significant performance gap between the model without any resampling technique and the best-performing approaches emphasizes how ignoring class imbalance leads to biased predictions, particularly disadvantaging the minority class. While ensemble methods alone provide only marginal improvements, integrating Resampling and Cost-Sensitive Learning proves to be the most effective way to improve the performance of the model in this task.
 \section{Discussion}
\label{sec:discussion}

\subsection{Detecting abandoning users}
Churn prediction has important social and practical implications~\citep{panimalar2023customer}. Here we defined and tackled the novel task of predicting churn (i.e., user abandonment) following a massive deplatforming moderation intervention on Reddit~\citep{cima2025investigating}. In consideration of the novelty of the task ---tackled here for the first time--- our results are promising. For example, we achieved \textit{micro F1} $= 0.914$  when detecting users who immediately halted activity after the moderation intervention (hard abandonment), and \textit{micro F1} $= 0.814$ when detecting users who initially maintained some activity but who eventually left Reddit (soft abandonment). Albeit preliminary, these results demonstrate that the task of preemptively estimating the effects of future moderation interventions is indeed feasible. Adding to the encouraging findings are the results of our leave-one-out cross-validation (LOOCV) analysis, which underscore the absence of substantial performance degradation when applying our model to unseen data and shows that the model learned generalizable behavioral patterns. At the same time however, our study highlighted some of the challenges of this new task. Among them are the difficulties at forecasting complex and heterogeneous user reactions~\citep{cresci2022personalized}, a dire class imbalance~\citep{robertson2022uncommon}, a partial and unreliable view of platform moderation actions~\citep{trujillo2023dsa}, and the growing difficulties at accessing platform data~\citep{jaursch2024dsa}, which culminate in the lack of extensive labeled datasets. The inherent challenges of predicting the effects of moderation interventions ---exemplified by our promising yet preliminary results--- imply that much work remains to be done before moderators and platform administrators will be able to leverage powerful and dependable tools for accurately estimating the effects of their actions. To this end, this study paves the way for future endeavors in this emerging area of content moderation, which proposes a new pathway for empowering moderators to optimize interventions based on desired outcomes, such as reducing toxicity while minimizing user attrition and churn. This is currently an as-yet unexplored frontier of moderation, which will open up once it achieves high performance in tasks such as the one proposed and tackled in this work.

\subsection{Characterizing moderated users}
Although the performances of the models developed in this study are promising, there is much room for improvement. Model performance strongly depends on the informativeness of the provided features. In this initial study, we computed and tested an extensive and diverse set of 142 features, many of which were borrowed from recent literature on related tasks~\citep{varol2018feature,habib2022proactive,tardelli2022detecting,mazza2022investigating}. Our set of features included information about the activity of the users before the moderation intervention, their participation in multiple communities, their relationships with other users, their use of toxic or otherwise aggressive speech, and their writing style. 
While some of the extracted features turned out to be powerful predictors in our models, a large share of the features that we evaluated was not particularly informative, which hindered model performance. This suggests that our task, proposed here for the first time, is substantially different from others previously addressed in related literature and for which those features provided robust performance. 
This finding underscores the importance of tailoring feature selection to the specific characteristics of the task, rather than relying on already-proposed features. Future efforts towards the prediction of moderation intervention effects should thus consider implementing and experimenting with a broader and even more diversified set of features.

Understanding the characteristics that influence user abandonment following a moderation intervention also provides valuable theoretical insights into online community dynamics. Our feature importance analysis highlights key behavioral patterns that distinguish users who leave the platform from those who remain, shedding light on possible psychological and sociological mechanisms underlying their decisions. As discussed in Section~\ref{sec:feat-imp}, the trend of comments in non-banned subreddits is among the most informative features. Users who were already decreasing their activity outside the banned communities before the intervention were more likely to abandon the platform entirely, suggesting that their engagement was closely tied to the restricted communities. This aligns with prior research suggesting that individuals with strong ties to specific groups may struggle to integrate into new ones after sudden disruptions~\citep{brown2000social,hu2024social}. Additionally, the ratio between the number of non-banned and banned subreddits a user participated in proved highly predictive of abandonment. Users who were more deeply embedded in banned communities, with relatively few interactions elsewhere, were more likely to leave. This supports the idea that having alternative engagement opportunities can act as a buffer against platform departure. Indicators of toxicity also emerged as strong predictors of abandonment. Users with higher toxicity levels were more likely to leave after a ban, potentially reflecting a pattern where users who frequently engage in disruptive behaviors feel alienated or unwilling to conform to platform norms following intervention. This observation aligns with theories in digital sociology that suggest enforcement actions can reinforce disengagement among users who strongly identify with rule-violating behaviors~\citep{thomas2021behavior}. These findings offer preliminary insights into how moderation interventions shape user behavior, with implications for both theoretical understanding and practical applications.

Based on these observations, promising directions for future work involve the development of socio-psychological features. In fact, predicting the effects of moderation interventions essentially revolves around predicting user behavior. Moreover, it is known that socio-psychological characteristics influence online behaviors, including toxic and aggressive ones~\citep{cresci2022personalized,giorgi2025human}. A recent demonstration of the latter can be found in those works who measured a significant improvement in hate speech detection tasks, when incorporating socio-psychological features~\citep{raut2023enhancing}. By the same token, those features could provide valuable information also in the related task of predicting behavioral changes following a moderation intervention, and particularly for those interventions targeting hateful or toxic users.

\begin{table}[t]
    \small
    \centering
    \setlength{\tabcolsep}{4pt}
    \begin{tabular}{clrrrrrr}
        \toprule
        & & & & \textbf{training time} (s) & \multicolumn{3}{c}{\textbf{test time} (s)} \\
        \cmidrule(r){5-5}\cmidrule(l){6-8}
        \textbf{task} & \textbf{model} & \textit{p} & \textit{F1} & \textit{total}&  \textit{features} & \textit{avg} & \textit{total} \\
        \midrule
        \multirow{7}{*}{\rotatebox[origin=c]{90}{\textbf{hard abandonment}}}
            & \texttt{KNN} & 20 & 0.384 & 0.003 & 1,138.809 &0.003 & 9.553 \\
            & \texttt{NB}  & 10 & 0.297 & 0.062 & 1,131.399 &0.001 & 4.696 \\
            & \texttt{RF}  & 142& 0.675 & 45.976 & 2,540.524 &0.018 & 57.311 \\
            & \texttt{AB}  & 142 & 0.659 & 10.648 & 2,540.524&0.011 & 37.132 \\
            & \texttt{DT}  & 10& 0.430 & 0.062 & 1,132.099 &0.001 & 3.982 \\
            & \texttt{GB}  & 142& 0.695 & 36.877 & 2,540.524 &0.005 & 16.483 \\
            & \texttt{SVM} & 50& 0.434 & 6.443 & 2,521.658 &0.003 & 9.195 \\
        \midrule
        \multirow{7}{*}{\rotatebox[origin=c]{90}{\textbf{soft abandonment}}}
            & \texttt{KNN} & 10& 0.473 & 0.021 & 1,101.728 &0.002 & 7.5051 \\
            & \texttt{NB}  & 10& 0.438 & 0.007 & 1,101.728 &0.002 & 4.948\\
            & \texttt{RF}  & 142 & 0.578 & 11.612  & 2,540.524 &0.007 & 24.189 \\
            & \texttt{AB}  & 142 & 0.588 & 11.693 & 2,540.524 &0.012 & 40.316
            \\
            & \texttt{DT}  & 20 & 0.494 & 0.234 & 1,136.813&0.002 & 5.469 \\
            & \texttt{GB}  & 142 & 0.590 & 23.668 & 2,540.524 &0.005 & 16.270 \\
            & \texttt{SVM} & 50 & 0.499 & 11.189 & 1,155.545&0.003 & 10.460 \\
        \bottomrule
    \end{tabular}
    \caption{\rev{Training and test times (in seconds) for each model across both abandonment tasks. Test performance is reported in terms of time to extract the features (\textit{features}), average prediction time per user (\textit{avg}), and total prediction time (\textit{total}). Since training is performed once on the full dataset, only the total training time (\textit{total}) is reported. Additionally, we report the number of features used (\textit{p}) and the resulting \textit{F1} score of each model.} }
    \label{tab:comp-cost}
\end{table}
 
\subsection{Scalability and Implementation}
For real-world deployment, scalability and computational efficiency are key considerations. Our approach leverages traditional machine learning models, which are computationally efficient and significantly less resource-intensive than deep learning-based alternatives. Additionally, many of our features can be precomputed and updated incrementally, reducing the overhead of real-time inference. Notably, our feature importance analysis indicates that only a subset of features is sufficient for accurate predictions, suggesting that models could be further optimized by focusing on the most informative features. 

\rev{To estimate the computational cost of our approach in a real-world scenario, we trained and tested our machine learning models on a training set of 12,994 users, a test set of 3,249 users, and a total of 1.57M comments. We measured both training and inference times for the two abandonment tasks. All experiments were conducted on a virtual machine running Ubuntu 20.04 with 32 GB of RAM and 16 virtual CPUs (Intel Xeon, 2.2 GHz). The results are reported in Table~\ref{tab:comp-cost}. As expected, across both tasks, simpler models exhibit the lowest training and test times, whereas the ensemble models require significantly more, up to several orders of magnitude in the worst cases. This is attributable to their greater complexity and the considerably higher number of features they use. Nonetheless, there is a significant trade-off in terms of F1 score, given that Gradient Boosting (\texttt{GB}), AdaBoost (\texttt{AB}), and Random Forest (\texttt{RF}) emerge as the best-performing models. The feature extraction time is significantly high for all models, including those using the lowest number of features. This is due to the reliance on a subset of computationally expensive features, such as part-of-speech tagging and toxicity scores, which dominate the overall extraction cost. However, in our specific case real-time processing is not a strict requirement as large-scale moderation interventions (e.g., subreddit bans) are relatively infrequent and do not require continuous, high-speed predictions. Therefore, it is feasible to favor more complex models in exchange for better predictive performance. Results reported in Table~\ref{tab:comp-cost} are also valuable to highlight the trade-off between model efficiency and effectiveness.} To conclude, future implementations could explore parallel computing, model compression techniques, or cloud-based deployment to further enhance efficiency.

\subsection{Cross-platform Applicability}
The application of our predictive approach to estimating the effects of moderation intervention across different online platforms presents both opportunities and challenges. In principle, our approach is general and its success in a new environment primarily depends on data availability and the ability to leverage platform-informative features. However, each platform has unique conventions, user behaviors, and moderation practices, which affect the generalizability of predictive models. Nonetheless, the existence of platforms featuring similar characteristics provide initial avenues for adaptation. For instance, Reddit shared structural and behavioral similarities with Voat, a now-defunct platform that attracted deplatformed communities. YouTube’s dynamics resemble those of BitChute, where content creators rely heavily on subscriber engagement. Likewise, Twitter and Weibo operate as fast-paced, influencer-driven networks, while TikTok and Instagram emphasize visual content and algorithm-driven exposure. These similarities could suggest that models trained on one platform could be transferred to analogous platforms with relatively minor modifications. At the same time however, major differences between platforms must be carefully considered. Platforms might differ not only in affordances and user interaction patterns, but also in their moderation policies~\citep{trujillo2023dsa}. While Reddit primarily enforces moderation at the community level (e.g., subreddit bans) ~\citep{trujillo2021echo,cima2025investigating}, Twitter/X and Facebook focus more on individual accounts ~\citep{pierri2023does,kalsnes2021hiding}, and YouTube's interventions often revolve around demonetization or content takedowns ~\citep{zappin2022youtube,ma2022m}. These differences imply that a system trained to estimate the effects of a certain type of intervention cannot be directly used to predict the effects of a different intervention, even within the same platform. Each intervention type ---whether account suspensions, content removals, or demonetization--- affects users in distinct ways, requiring models specifically tailored to the nature of the intervention. Ultimately, the heterogeneous landscape of online platforms and their evolving moderation practices pose challenges to cross-platform applicability. Adapting predictive moderation tools requires both scientific and practical efforts to ensure that models remain accurate and relevant in different digital ecosystems. Future research should explore the trade-off between platform-independent and platform-specific features so as to best capture user engagement and retention dynamics, while also considering how shifting moderation strategies affect the accuracy of predictive models.

\subsection{Predicting effects: classification, quantification, regression}
In this work we tackled the problem of predicting the users that would abandon Reddit following a moderation intervention as a binary classification task. Currently, there is a dire lack of tools and systems to predict or estimate the effects of moderation interventions. Within this context, our work paves the way to the development of decision support tools to assist moderators and platform administrators in carrying out effective and efficient content moderation, thus overcoming the limitations of the current trial-and-error approach~\citep{trujillo2022make}.
In fact, in light of the promising results that we achieved, we envision the possibility to develop additional tools in the near future, to further support moderation endeavors by leveraging established machine learning paradigms. Among them are quantification and regression approaches. For instance, regarding the former, a platform may not be interested in knowing which specific users are likely to abandon or remain on the platform following a moderation intervention. Such a need could arise, among other reasons, due to privacy concerns~\citep{gorwa2020algorithmic}. In that case, effective moderation could still be guaranteed by computing aggregated estimates of how many users would likely leave compared to those who would remain. From the methodological standpoint, such a problem would be better addressed as a \textit{quantification} task rather than as a classification one~\citep{gonzalez2017review}. As a matter of fact, recent results on quantification have demonstrated the superiority of this approach over \textit{classify-and-count} strategies~\citep{esuli2023learning}. Quantification tasks for obtaining aggregated estimates of the effects of moderation interventions thus represent a favorable avenue for future research. 
Another limitation of classification approaches is the narrow view it offers on post-moderation user behavior. As an example, in order to tackle the binary classification task, we had to come up with a definition for abandoning users. Given the inherent complexity and multifaceted nature of user behavior, we experimented with two different definitions of abandonment ---\textit{hard} and \textit{soft}, each with its own strengths and weaknesses--- so as to account for a broader spectrum of heterogeneous user behavior. Nonetheless, a binary classification task inevitably conceals important details about actual user reactions~\citep{trujillo2023one}. For this reason, we envision the possibility to cast the prediction of the effects of moderation interventions as a \textit{regression} task~\citep{tardelli2022detecting}. This nuanced approach would avoid the need for limiting binary labels in favor of fine-grained estimates of the expected behavioral changes. By moving beyond simplistic labels, such an approach would empower platform moderators with deeper insights into user reactions and patterns, facilitating targeted and effective intervention strategies to foster user retention and community growth. However, solving a regression task with sufficient accuracy poses even more challenges than those addressed here. To this end, additional work on feature engineering is essential to provide the models with more informative features. Moreover, further work should also be directed towards model development, for example by employing sophisticated models capable of fully exploiting the provided features.

\subsection{Ethical considerations}
\label{sub:ethical}
The use of predictive models in content moderation may raise ethical concerns regarding privacy, bias, and transparency. While our study relies solely on publicly available and anonymized Reddit data, real-world applications must ensure compliance with data protection regulations and implement strict privacy safeguards to avoid over-surveillance of users. Biased predictions are another concern. To mitigate bias, we applied feature selection and rebalancing techniques to prevent skewed predictions. Nonetheless, predictive models can still reflect underlying biases in moderation histories. Future research should audit platform moderation practices for fairness, so as to minimize possible unintended discrimination, and explore fairness-aware machine learning approaches for making predictions. Additionally, the inclusion of potentially sensitive user attributes as features should be carefully weighed against their actual predictive value, ensuring that their use is minimized or avoided whenever possible to uphold fairness and privacy. Transparency is also crucial for maintaining trust in predictive moderation. While our study employs SHAP and PI to analyze feature importance, real-world implementations should prioritize explainable AI methods to ensure even better interpretability. Finally, predictive models should not replace human decision-making but serve as tools to support moderators, ensuring that intervention strategies remain fair and accountable.

\subsection{Limitations}
\label{sub:limitations}
This study focuses on a set of Reddit users who experienced a specific form of moderation intervention (i.e., multiple community bans) at a specific time, which may limit the generalizability of our findings. Each platform operates within a unique ecosystem with its own user demographics, interaction norms, and moderation policies. Moreover, user behavior and platform moderation dynamics may change and evolve through time. Therefore, in spite of our efforts at estimating the generalizability of our results, findings derived from a single platform and for a single intervention may not be fully applicable to other platforms or interventions. Furthermore, we focused on a subset of all moderated users: those who were particularly active in the banned communities. As such, our models might exhibit decreased performance if applied to users with markedly different characteristics. Additionally, the specific definitions of hard and soft abandonment that we proposed here might represent a further limitation, given that alternative definitions may yield different interpretations of user behavior and corresponding ground truths for the machine learning tasks. Another limitation of our work stems from the set of features with which we experimented. Despite implementing a relatively large number of features that convey diverse information, many important dimensions of user behavior remain unexplored for this task. Encoding those dimensions in effective machine learning features would boost the performance of the trained models, allowing better results. By the same token, in this first study on predicting the effect of a moderation intervention, we focused on detecting abandoning users. However, effects of a moderation intervention can be defined and investigated in multiple other ways that do not necessarily involve user abandonment, but rather the degree of toxic speech on the platform after the intervention, the extent of polarization, the reliance on factual news, and more. The predictability of these and other moderation effects remains to be assessed.
 \section{Conclusions and Future Work}
\label{sec:conclusions}
We proposed and tackled ---for the first time--- the task of predicting the effects of a moderation intervention. Specifically, we detected those users who abandoned Reddit following the ban of a large number of communities on the platform. To solve this task, we investigated the behavior of 16,540 users by leveraging 16 million comments they posted in the banned communities and 13.8 million comments they posted in non-banned ones. Starting from this extensive dataset, we extracted 142 features conveying information about the activity, toxicity, relations, and writing style of the analyzed users. Our results are promising, albeit preliminary as one would expect from a new task, with the best model achieving \textit{micro F1} $= 0.914$. Our model proved to be sufficiently generalizable when applied to users from unseen communities. Furthermore, we found that activity features are the most informative for this task, followed by relational and toxicity features. Conversely, writing style features seem to provide limited information.

Given the novelty of the task, promising directions for future work are multifold. Among them are the inclusion of additional features that could provide relevant information on user behavior, and particularly, on their possible reactions to a moderation intervention, such as socio-psychological features. Other promising features are those that allow tracking linguistic and semantic shifts over time, including features based on sentiment, topics, and linguistic diversity. Future research could also focus on improving generalizability through the adoption of new datasets, enhancing scalability, enabling real-time applications, and ensuring multi-platform compatibility. Additionally, studying longitudinal changes in user behavior and their effects on abandonment predictions over time could provide deeper insights into the long-term impact of moderation interventions. Other favorable avenues of experimentation involve the adoption of more powerful machine learning methods, such as deep learning-based techniques (e.g., LSTMs or transformers). Finally, the task of predicting the effects of a moderation intervention could be cast as a multi-class, regression, or quantification problem, rather than as a binary classification as done in the present work. In this regard, being able to precisely estimate the extent of behavioral change exhibited by affected users would provide even more fine-grained information to platform administrators for planning their moderation actions.
 
\section*{Acknowledgments}
This work is partially supported by the European Union -- Next Generation EU, Mission 4 Component 1, for project PIANO (CUP B53D23013290006); by the PNRR-M4C2 (PE00000013) "FAIR-Future Artificial Intelligence Research" - Spoke 1 "Human-centered AI", funded under the NextGeneration EU program; and by the Italian Ministry of Education and Research (MUR) in the framework of the \texttt{FoReLab} project (Departments of Excellence).

\printcredits

\bibliographystyle{unsrt}

\bibliography{references}

\end{document}